\documentclass[apj,onecolumn,tighten]{emulateapj}
\usepackage{mathrsfs}
\usepackage{amsmath}
\usepackage{color}
\usepackage{xcolor}

\begin{document}

\title{{\sc CosmosDG}: An {\it hp}-Adaptive Discontinuous Galerkin Code for Hyper-Resolved Relativistic MHD}

\author{Peter Anninos}
\affil{Lawrence Livermore National Laboratory \\
       P.O. Box 808, Livermore, CA 94550, USA}
\author{Colton Bryant}
\affil{Department of Engineering Sciences \& Applied Mathematics \\
       Northwestern University, 2145 Sheridan Rd, Evanston, Illinois, 60208, USA}
\author{P. Chris Fragile}
\affil{Department of Physics \& Astronomy \\
       College of Charleston, 66 George Street, Charleston, SC 29424, USA}
\author{A. Miguel Holgado}
\affil{Department of Astronomy \& National Center for Supercomputing Applications \\
       University of Illinois at Urbana-Champaign, Urbana, Illinois, 61801, USA}
\author{Cheuk Lau}
\affil{Lawrence Livermore National Laboratory \\
       P.O. Box 808, Livermore, CA 94550, USA}
\author{Daniel Nemergut}
\affil{Operations \& Engineering Division \\
       Space Telescope Science Institute, 3700 San Martin Drive, Baltimore, MD 21218, USA}

\begin{abstract}
We have extended {\sc Cosmos++}, a multi-dimensional unstructured adaptive mesh code for solving the 
covariant Newtonian and
general relativistic radiation magnetohydrodynamic (MHD) equations, to accommodate both discrete
finite volume and arbitrarily high order finite element structures. The new finite element implementation, called {\sc CosmosDG},
is based on a discontinuous Galerkin (DG) formulation, using both 
entropy-based artificial viscosity and slope limiting
procedures for regularization of shocks. High order multi-stage forward
Euler and strong stability preserving Runge-Kutta
time integration options complement high order spatial
discretization. We have also added flexibility in the code infrastructure allowing
for both adaptive mesh and adaptive basis order refinement to be performed separately or
simultaneously in a local (cell-by-cell) manner.
We discuss in this report the DG formulation 
and present tests demonstrating the robustness, accuracy, and convergence of our numerical
methods applied to special and general relativistic MHD, though we note an equivalent capability
currently also exists in {\sc CosmosDG} for Newtonian systems.
\end{abstract}
\maketitle

\section{Introduction}
\label{sec:intro}

Discontinuous Galerkin (DG) finite element (FE) methods have raised great interest
over the past few decades,
particularly in the engineering communities and applied mathematics literature
\citep[e.g.][]{Johnson84,Cockburn89,Kershaw95,Cockburn98,Hartmann02,Kuzmin04,Guermond11}.
These methods were originally introduced more than forty years ago
for neutron transport
\citep{Reed73}, but they
have since expanded in scope and become popular for solving
more general systems of conservation laws across a variety
of physical disciplines, including computational 
fluid dynamics, acoustics, and electromagnetics.
However, with the exception of a few groups, they have 
yet to be widely adopted in computational astrophysics, particularly
relativistic astrophysics, where finite difference
and finite volume (FV) methods dominate.
One interesting attempt to use it was presented in \citet{Meier99}.  
In that work, the Einstein field equations were discretized in all four spacetime dimensions, 
thus treating time entirely equivalent to space. We are aware of only three
other applications of DG to relativistic magnetohydrodynamics (MHD): \citet{Radice11},
\citet{Zanotti15} and \citet{Kidder17}. The first two have significant restrictions, with the first being limited to one-dimensional spherical symmetry, while the second is limited to special relativity. The
third paper considers multi-dimensional relativistic MHD, though without adaptivity.
The methods described in the current paper apply to both special and general relativistic
MHD, to multi-dimensional spacetimes with no symmetry restrictions, and
to dynamically adaptive mesh and polynomial representations.

Finite element methods possess
a number of desirable properties, including their suitability for conservation equations,
their cost competitiveness with finite volume methods, their compatibility with Riemann solvers,
their potential for achieving spectral-like convergence rates,
and their applicability to
unstructured meshes and local (cell-by-cell) refinement.  
They are related to FV methods in the sense that basic
cell-centered FV schemes correspond to the DG(1) method, i.e., to the
discontinuous Galerkin method using piecewise linear polynomials ($p=1$). 
Consequently, the DG($p$) method, with $p > 1$, can be regarded as a natural 
extension of the FV method to higher orders.  
For continuous FE methods, however, high order comes at a cost as it requires the 
storage and inversion of a very large, global matrix.  Such global matrix inversions 
do not scale well to large numbers of processors and, therefore, limit the sizes and 
types of problems to which continuous FE can be applied.  
This was the shortcoming of the \citet{Meier99} implementation.  In contrast,
the discontinuous nature of DG methods allows high-order polynomial approximations 
to be made within a single element rather than across wide stencils, as in the 
case of high-order FV methods, or the entire grid, as in continuous FE methods.  
Thus, all matrix inversions are done locally, rather than globally.  Furthermore, 
as elements only need to communicate with adjacent elements with a common 
face (von Neumann neighbors), regardless of the order of accuracy of the scheme, 
inter-element communications are minimal, making the method highly parallelizable. 
Furthermore, the parallelization can be efficiently accomplished through simple domain decomposition.

Another advantage of DG methods is that they are relatively straightforward to implement on 
unstructured meshes.  Unstructured meshes, themselves, have the advantages that 
they can more accurately discretize complex geometries,
easily adapt to surface boundaries, and 
enhance solution accuracy and efficiency through the use of {\em local} (i.e.
cell-by-cell) adaptive mesh refinement (AMR, also commonly referred to
as $h$-refinement).
The discontinuous nature of DG methods relaxes the strong continuity 
restrictions of continuous FE methods, leaving elements free to be 
refined or coarsened, without affecting solutions or data structures in other elements. 

DG methods also allow easy implementation of $p$-refinement, 
or adaptive order refinement (AOR), where the polynomial 
degree of the basis is varied.  Thus, the order of accuracy can be different 
from element to element.  This type of refinement is potentially very powerful, 
as exponential convergence rates are possible when solutions are smooth \citep{Babuska86,Schwab99}.  
Combined with $h$-refinement, these high rates of convergence are even possible 
when singularities are present \citep{Schwab99}.

For all of these reasons DG methods are particularly appealing and a natural
progression for {\sc Cosmos++} \citep{Anninos03b,Anninos03a,Anninos05}, 
an unstructured $hr$-adaptive mesh code we have developed for
both Newtonian and general relativistic astrophysical applications.
So we have taken this opportunity to upgrade {\sc Cosmos++} for
both FV and DG frameworks, and to implement $p$-refinement.
{\sc Cosmos++} supports numerous physics packages, including hydrodynamics,
ideal magnetic fields, primordial chemistry, nuclear reaction networks,
Newtonian self-gravity, dynamical general relativistic spacetimes, 
radiation transport, molecular viscosity, thermal conduction, etc., 
but we focus exclusively on the
MHD in this paper since only those packages have been
generalized so far to work within {\sc CosmosDG}.
The equations, methods, and tests of our code are described in the remaining sections, 
emphasizing the DG aspects.
The reader is referred to our previous papers, most notably \citet{Anninos05},
for additional details not included in this paper, such as mesh hierarchy constructions, parallelism,
and class inheritance designs.
Unless otherwise noted, standard index notation is used for
labeling spacetime coordinates: repeated indices represent summations,
raising and lowering of indices is done with the 4-metric tensor, and
Latin (Greek) indices run over spatial (4-space) dimensions.

\section{Basic Equations}
\label{sec:equations}

\subsection{General Relativistic MHD}

We begin by writing the contravariant stress energy density
tensor for a viscous fluid with ideal MHD as a linear combination of the
hydrodynamic, magnetic and viscosity contributions:
\begin{equation}
T^{\alpha\beta} =
         (\rho h + 2 P_B/c^2 - Q_B/c^2) u^\alpha u^\beta 
         + (P+P_B-Q_B) g^{\alpha\beta} 
         - Q^{\alpha\beta}_S - b^\alpha b^\beta ~.
\label{eqn:gr_tmn}
\end{equation}
Here
$\rho$ is the fluid mass density,
$h=1+\epsilon/c^2 + P/(\rho c^2)$ is the specific enthalpy,
$c$ is the speed of light,
$u^\alpha = u^0 V^\alpha$ is the contravariant velocity,
$V^\alpha$ is the transport velocity,
$P$ is the fluid pressure
(for an ideal gas $P=(\Gamma-1)e$, $e=\rho\epsilon$ is the fluid internal energy
density, $\Gamma$ is the adiabatic index), $b^\alpha$ is the magnetic field,
$P_B = g_{\alpha\beta}b^\alpha b^\beta/2$ is the magnetic pressure,
$Q_B$ is bulk viscosity,
$Q^{\alpha\beta}_S$ is the symmetric shear viscosity tensor representing artificial
or molecular viscosity,
and $g_{\alpha\beta}$ is the curvature metric.
Although {\sc Cosmos++} supports molecular viscosity, it is not currently
incorporated into the DG framework, so we do not consider it further in this paper.

The four fluid equations
(energy and three components of momentum) are derived
from the conservation of stress energy:
$\nabla_\mu T^\mu_{\ \nu} = \partial_\mu T^\mu_{\ \nu} + 
 \Gamma^\mu_{\alpha\mu} T^\alpha_{\ \nu} - \Gamma^\alpha_{\mu\nu} T^\mu_{\ \alpha} = S_\nu$,
where $\Gamma^\alpha_{\mu\nu}$ are the Christoffel symbols and $S_\nu$
represent arbitrary source terms.
In addition to energy and momentum, we also require equations
for the conservation of mass
$\nabla_\mu(\rho u^\mu) = \partial_t(\sqrt{-g} u^0\rho) + \partial_i (\sqrt{-g} u^0 \rho V^i) = 0$,
and magnetic induction $\nabla_\mu(u^\mu b^\nu - b^\mu u^\nu) = 0$.

Expanding out the space and time coordinates, the four-divergence 
($\nabla_\mu T^\mu_\nu = S_\nu$) of the
mixed index stress tensor is written
\begin{equation}
\partial_t(\sqrt{-g} T^0_\nu) + \partial_i(\sqrt{-g} T^i_\nu)
    = \sqrt{-g} T^\mu_\sigma \ \Gamma^\sigma_{\mu\nu} + \sqrt{-g} S_\nu ~.
\end{equation}
Further defining energy and momentum as
${\cal E} = - \sqrt{-g} {T}^0_0$ and
${\cal S}_j = \sqrt{-g} {T}^0_j$, the equations take on a traditional transport formulation
\begin{equation}
\partial_t {\cal E} + \partial_i( - \sqrt{-g} T^i_0)
    = - \sqrt{-g} T^\mu_\sigma \ \Gamma^\sigma_{\mu 0} ~.
\end{equation}
\begin{equation}
\partial_t {\cal E} + \partial_i({\cal E} V^i)
    + \partial_i ( \sqrt{-g} (P + P_B) V^i)
    = - \sqrt{-g} T^\mu_\sigma \ \Gamma^\sigma_{\mu 0} ~,
\end{equation}
for energy, and
\begin{equation}
\partial_t {\cal S}_j + \partial_i( \sqrt{-g} T^i_j)
    = \sqrt{-g} T^\mu_\sigma \ \Gamma^\sigma_{\mu j} ~,
\end{equation}
\begin{equation}
\partial_t {\cal S}_j + \partial_i({\cal S}_j V^i)
    + \partial_i ( \sqrt{-g} (P + P_B)~g^0_j~V^i)
    = \sqrt{-g} T^\mu_\sigma \ \Gamma^\sigma_{\mu j} ~.
\end{equation}
for momentum. Completing the system of equations, 
energy and momentum conservation are supplemented with mass conservation
\begin{equation}
\partial_t D + \partial_i (D V^i) = 0 ~,
\end{equation}
where $D=\sqrt{-g} u^0 \rho = W\rho$ is the boost density, 
and magnetic induction
\begin{equation}
 \partial_t B^j + \partial_i(B^j V^i - B^i V^j) =
       \eta g^{ij} \partial_i (\partial_k B^k) ~,
      \label{eqn:mag}
\end{equation}
where $B^\mu = W(b^\mu - b^0 V^\mu)$ (with $b^0=-B^\alpha V_\alpha/(W V^\alpha V_\alpha)
= B^\alpha V_\alpha u^0/\sqrt{-g}$) is the evolved spatial ($B^0=0$), divergence-free 
($\partial_k B^k=0$) representation of the field, distinct from the rest frame field $b^\mu$.
The additional source term on the right hand side of equation (\ref{eqn:mag})
is a form of divergence cleanser used to drive equation (\ref{eqn:mag}) to
satisfy $\partial_k B^k=0$ on a scale defined by the choice of parameter $\eta$,
typically set proportional to the largest characteristic speed in the flow.

Mesh motion is easily accommodated by a straight-forward 
replacement of generic advective terms
\begin{equation}
\partial_t (\sqrt{-g} T^0_\alpha) + \partial_i(\sqrt{-g} T^0_\alpha V^i)
\end{equation}
with
\begin{equation}
\partial_t (\sqrt{-g} T^0_\alpha) + \partial_i(\sqrt{-g} T^0_\alpha (V^i - V_g^i)) + \sqrt{-g} T^0_\alpha \partial_i V_g^i ~,
\end{equation}
where $V_g^i$ is the grid velocity, and $T^0_\alpha$ is used here to represent any evolved field,
including $E$, $S_j$, $D$, and $B^j$.

\subsection{Newtonian MHD}

For comparison and future reference, we add in this section the equivalent covariant
form of the corresponding Newtonian MHD equations.
Drawing an analogy with the relativistic equations presented above,
we write the effective Newtonian stress energy tensor as
\begin{equation}
{\cal T}^{\alpha\beta} =
         \rho v^\alpha v^\beta + (P+P_B-Q_B) g^{\alpha\beta} 
         - Q_S^{\alpha\beta} - b^\alpha b^\beta ~.
\label{eqn:newt_tmn}
\end{equation}
In moving curvilinear coordinates the Newtonian conservation equations take
on a similar form as their relativistic counterparts
\begin{equation}
 \partial_t (\sqrt{g} \rho) +
  \sqrt{g} \rho \partial_i V_g^i +
 \partial_i \left(\sqrt{g} \rho (v^i - V_g^i)\right) = 0 ~,
 \label{eqn:newt_den}
\end{equation}
\begin{equation}
 \partial_t (\sqrt{g} E) +
  \sqrt{g} E \partial_i V_g^i +
 \partial_i \left(\sqrt{g} E (v^i - V_g^i) + \sqrt{g} F^{0i}\right) =
      - \sqrt{g} \rho v^i \partial_i \phi ~,
      \label{eqn:newt_ene}
\end{equation}
\begin{eqnarray}
 \partial_t( \sqrt{g} s_j) + 
  \sqrt{g} s_j \partial_i V_g^i +
 \partial_i \left(\sqrt{g} s_j (v^i - V_g^i) + \sqrt{g} F^i_{\ j}\right) =
        \sqrt{g} {\cal T}^{i k} \Gamma_{i k j}
       -\sqrt{g} \rho \partial_j \phi ~,
    \label{eqn:newt_mom_dn}
\end{eqnarray}
\begin{equation}
 \partial_t (\sqrt{g} b^j) +
  \sqrt{g} b^j \partial_i V_g^i +
 \partial_i \left(\sqrt{g} b^j (v^i - V_g^i) - \sqrt{g} b^i v^j\right) =
       \eta g^{ij} \partial_i (\partial_k (\sqrt{g} b^k)) ~.
      \label{eqn:newt_mag}
\end{equation}
with flux terms
\begin{equation}
  F^{0i}     = (P+P_B) v^i - b^i b_j v^j ~,
\end{equation}
\begin{equation}
  F^{ij} = (P+P_B) g^{ij} - b^i b^j ~.
\end{equation}
Here $E$ is the total energy density
including internal, magnetic and kinetic energy contributions:
$E=e + b^i b_i/2 + \rho v^i v_i/2$.
This definition does not include gravitational energy which is
treated as an add-on (nonconservative) source represented by the
potential $\phi$ in the right-hand-sides of the energy and momentum equations.
In this form, Newtonian and relativistic fluxes and source terms are
easily interchangeable in the numerical solver frameworks.

\subsection{Primitive Fields}

At the beginning (or end) of each time cycle a series of coupled nonlinear equations
are solved to extract primitive fields (mass density, internal energy, velocity)
from evolved conserved fields (boost density, total energy, momentum), after which
the equation of state is applied to compute thermodynamic quantities like
pressure, sound speed and temperature.
For Newtonian systems this procedure is straightforward, but relativity
complicates the inter-dependency of primitives, 
and their extraction from conserved fields requires special iterative treatment.
We have implemented several procedures for doing this, solving
one, two, or five dimensional inversion schemes \citep{Noble06,Fragile12},
or a nine dimensional fully implicit method (including coupling terms)
when radiation fields are present \citep{Fragile14}.

One of the more robust procedures reduces the number of equations from the
number of evolved fields (five in the simplest case of hydrodynamics)
to two, taking advantage of projected conserved constraints
to facilitate the reduction.
This 2D method solves two constraints, energy and momentum,
derived from a projection of the stress energy tensor to the
normal observer frame with
four-velocity $n_\nu = [-\alpha, 0, 0, 0]$ and lapse function $\alpha$
\begin{equation}
\frac{\tau_\mu}{\alpha} = -\frac{n_\nu T^\nu_\mu}{\alpha} = T^0_\mu
       = (\rho h + 2 P_B) u^0 u_\mu 
         + (P + P_B) g^0_\mu - b^0 b_\mu ~.
\end{equation}
Defining
$\widetilde{B}^\mu = \alpha B^\mu/\sqrt{-g}$, the energy ${\cal E}$ and
momentum $\widetilde{m}^2$ constraints take the form
\begin{equation}
{\cal E} =  \tau_\mu n^\mu
\alpha T^0_\mu n^\mu =
    -\frac{\widetilde{B}^2 (1+v^2)}{2}
    +\frac{(\alpha T^0_\mu \widetilde{B}^\mu)^2}{2 w^2}
    - w + P ~,
\label{eqn:primenergy}
\end{equation}
\begin{equation}
\widetilde{m}^2 = \widetilde{\tau}^\mu \widetilde{\tau}_\mu
    =  \tau^{\mu} \tau_{\mu}
       + \alpha^2 (\tau^0)^2
    =  v^2 (w + \widetilde{B}^2)^2
       - \frac{(2w + \widetilde{B}^2)}{w^2} \left(
         \alpha T^0_\mu \widetilde{B}^\mu \right)^2 ~,
\label{eqn:primmomentum}
\end{equation}
where 
$\widetilde{\tau}^\nu = (g^\nu_\mu + n^\nu n_\mu) \tau^{\mu}$,
$v^2 = 1- (1/\gamma^2)$, $\gamma = \alpha u^0$ is the Lorentz boost,
$w = (\alpha u^0)^2(\rho h)$ is the scaled enthalpy, 
and the pressure and its gradients ($\partial P/\partial w$,
$\partial P/\partial v^2$) are calculated from the
ideal gas law
\begin{equation}
P = \frac{\Gamma-1}{\Gamma} \left( w(1-v^2) - {\alpha u^0 \rho}\sqrt{1-v^2} \right) ~.
\end{equation}
These constraints represent nonlinear equations
for the two unknowns, $w$ and $v^2$, and are solved
by Newton iteration. All of the other terms in these
equations are easily derived from evolved quantities.

An alternative, though generally more costly, option utilizes
Newton iteration to solve a full, unprojected matrix system of nonlinear
equations constructed from
the primitive field dependency of all of the conserved
or evolved quantities. Thus within each iteration one
constructs a ($5\times5$ for the case of hydrodynamics) Jacobian
matrix $A_{ij} = \partial U^i/\partial P^j$ evaluated at guess primitive
solutions. Here $U^i \equiv [D, E, S_k] = \sqrt{-g} [u^0\rho, - T^0_0, T^0_i]$
is a vector list of conserved fields, and
$P^j \equiv [\rho, \epsilon, \widetilde{u}^k]$
is a vector list of corresponding primitive fields.
We use $\widetilde{u}^k = u^k - u^0 g^{0k}/g^{00}$ with
$u^0=\gamma/\alpha$ as the primitive velocity in place of $v^2$
in this procedure.

\section{Numerical Methods}

\subsection{DG Framework}

The DG framework reviewed here is presented in the context of generic
conservation laws expressed in the following vector form:
\begin{equation}
\label{eqn:general_transport}
\partial_t u + \vec{\nabla} \cdot \vec{F}\left(u\right) = s 	~,
\end{equation}
where $u$ is the conserved quantity of interest (density, momentum, energy), 
$\vec{F}$ is the flux,  and $s$ is an arbitrary source term.
We switch to vector notation in this section in order not to confuse spacetime
indices with basis function labels or indexing of matrix elements.

We begin by multiplying equation 
(\ref{eqn:general_transport}) by a set of weight functions $p_i(x)$, and integrating the 
resulting equations over the volume ${\cal V}_k$ of each cell $k$
\begin{equation}
\label{eqn:weak_form_1}
\int_{{\cal V}_k} d{\cal V}~p_{i}(x)~\left( \partial_t u + \vec{\nabla} \cdot \vec{F}\left( u \right) \right) 
     = \int_{{\cal V}_k} d{\cal V}~p_{i}(x)~s	~.
\end{equation}
Although we have written the DG framework in a modular way, anticipating adding more options
for basis sets in the future, we have for this work adopted Lagrange interpolatory
polynomials defined as
\begin{equation}
p_i(x) = \prod_{k=1,k\neq i}^{n} \frac{x-x_k}{x_i-x_k} ~.
\end{equation}
The shape functions of this basis are unity at their respective nodes and zero at all other nodes.
A multi-dimensional version is constructed through tensor products of one-dimensional
polynomials on a unit reference element covered with
$(p+1)^n$ nodes, where $p$ is the order and $n$ is the number of dimensions.

The divergence theorem is then applied to
equation (\ref{eqn:weak_form_1}) which results in the so-called weak 
form of equation (\ref{eqn:general_transport})
\begin{equation}
\label{eqn:weak_form_2}
\int_{{\cal V}_k} d{\cal V}~p_{i} \partial_t u - 
\int_{{\cal V}_k} d{\cal V}~\vec{F}(u) \vec{\nabla} p_{i} + 
\int_{\partial k} dA~p_{i} \vec{n} \cdot \vec{H}(u^+,u^-) 
     = \int_{{\cal V}_k} d{\cal V}~p_{i} s	~,
\end{equation}
where $\partial k$ is the surface of cell $k$, $\vec{n}$ is the 
outward pointing vector normal to the surface,
and $\vec{H}(u^+,u^-)$ is an appropriately calculated flux at the cell boundaries.
$\vec{H}(u^+,u^-)$ takes into account discontinuities across cell faces, and 
depends on both interior and adjoining neighbor state solutions.

A simple method for determining
surface fluxes is standard upwinding, which uses the value of $u$ 
inside the cell for the exiting flux 
($\vec{H}(u^-) = \vec{F}(u(r_S - \epsilon \vec{n}))$ for $\vec{n} \cdot \vec{v} \geq 0$),
and the value outside for the incoming flux 
($\vec{H}(u^+) = \vec{F}(u(r_S + \epsilon \vec{n}))$ for $\vec{n} \cdot \vec{v} < 0$),
where $r_S$ is a location on the cell surface, and $\epsilon$ is some arbitrarily small positive value.
Alternative, less diffusive options for computing surface fluxes can be easily substituted
for simple upwinding. Our implementation currently supports both Lax-Friedrichs (LF)
and Harten-Lax-vanLeer (HLL) approximate Riemann solvers \citep{Harten83}
\begin{equation}
H_{LF}(u^+,u^-) = \frac{1}{2}\left(F(u^+) + F(u^-) - \alpha_+ (u^+ - u^-)\right)  ~,
\end{equation}
\begin{equation}
H_{HLL}(u^+,u^-) = \frac{1}{\alpha_+ + \alpha_-} \left(
                            \alpha_+ F(u^-) + \alpha_- F(u^+) - 
                            (\alpha_+ \alpha_-)(u^+ - u^-) \right)  ~,
\end{equation}
where $\alpha_\pm$ are the minimum and maximum characteristic wave speeds.
The relation of DG methods to Riemann solvers thus comes from the discontinuous 
representation of the solution at element interfaces, which requires a relaxation of the 
cross-element continuity condition.  Instead of enforcing a single, continuous 
solution at element interfaces, like the continuous FE method, the DG method 
supports ``left'' and ``right'' states on either side of the interface similar
to finite volume or finite difference methods.  It then 
treats the element boundary by solving a local Riemann problem to calculate the 
appropriate flux, ensuring the method remains conservative while
capturing the shock characteristics.

Time and space dependencies
of each evolved quantity and source term are split into separable form and
expanded using a set of spatial basis functions, which for Galerkin
methods are equal to the weight functions $p_j(x)$ introduced earlier
\begin{equation}
\label{eqn:expansion}
u = \sum_{j=1}^J u_{j}(t)~p_{j}(x) 	~.
\end{equation}
Substituting equation (\ref{eqn:expansion}) into (\ref{eqn:weak_form_2})
and performing the integrals with quadratures produces the following
linear system for the expansion (or support) coefficients ${\bf u} \equiv u_j(t)$:
\begin{equation}
\label{eqn:weak_form_3}
{\bf M} \partial_t {\bf u} - {\bf S} {\bf u} + {\bf R} {\bf u} = {\bf M} {\bf s} ~.
\end{equation}
${\bf M}$ is the mass matrix associated with each cell
\begin{equation}
\label{eqn:mass_matrix}
M_{ij} = \sum_{q=1}^{Q_V}p_{iq}p_{jq}w_{q}	~,
\end{equation}
where $Q_V$ is the number of quadrature points for the volumetric integral 
of cell $k$, $p_{iq}$ is the $i$-th weight function evaluated at the $q$-th 
quadrature point, and $w_q$ is the weight of the $q$-th 
quadrature point. ${\bf S}$ is the stiffness matrix defined as
\begin{equation}
\label{eqn:stiffness_matrix}
S_{ij} = \sum_{q = 1}^{Q_V} p_{jq} \vec{v}_{q} \cdot \vec{\nabla} p_{iq} w_{q},
\end{equation}
where $\vec{v}_{q}$ is the velocity at the $q$-th quadrature point within the cell volume. 
${\bf R}$ is the surface matrix defined for each element as
\begin{equation}
\label{eqn:surface_matrix}
R_{ij} = \sum_{q=1}^{Q_S}p_{jq}\vec{v}_{q}p_{iq}w_{q}	~,
\end{equation}
where $Q_S$ is the number of quadrature points for the integral over the 
surface (faces) of each cell. 
In general, the weight functions are defined over a 
reference cell with local coordinates $\xi_i$, and mapped onto each physical cell 
with global coordinates $\eta_j$ using a Jacobian matrix {\bf J} defined for each 
element as
\begin{equation}
\label{eqn:jacobian_matrix}
J_{ij} = \sum_{m = 1}^{M} \left(\partial_{\xi_i} p_m \right) \eta_{jm}  ~,
\end{equation}
where $M$ is the number of weight functions, $p_m$ is the $m$-th weight function 
defined over the reference element, and $\eta_{jm}$ is the location of the $m$-th support
point with respect to the $j$-th global coordinate. 
To apply the Jacobian mapping, we multiply equations (\ref{eqn:mass_matrix}), 
(\ref{eqn:stiffness_matrix}), and (\ref{eqn:surface_matrix}) by the determinant of {\bf J},
which is built separately for each zone and cell face elements.

The DG formulation is completed with a procedure for discretizing
the remaining time derivative term in equation (\ref{eqn:weak_form_3}). 
{\sc Cosmos++} and {\sc CosmosDG} support numerous high order time integration options, some of which are
discussed below in section \ref{subsec:timeint}, but we write out an explicit expression for illustration 
here using a simple, single-step forward Euler solution
\begin{equation}
\label{eqn:weak_form_4}
{\bf u}^{n + 1} = {\bf u}^n + \Delta t^n \left({\bf M}\right)^{-1} 
            \left({\bf S}^n {\bf u}^n - {\bf R}^n {\bf u}^n + {\bf M} {\bf s}^n \right)	~,
\end{equation}
where $n$ denotes the time level, and $\Delta t^n$ is the time step size.
Equation (\ref{eqn:weak_form_4}) can alternatively be written as
\begin{equation}
\label{eqn:weak_form_5}
{\bf u}^{n + 1} = {\bf u}^n + \Delta t^n \left( ({\bf M})^{-1} {\bf B}^n + {\bf s}^n \right) ~,
\end{equation}
where we have absorbed the evolved and velocity fields into an inclusive
flux term and merged the two source matrices into a single source combining
volume and surface contributions
\begin{equation}
B_{ij} = 
          \sum_{q = 1}^{Q_V} p_{jq} \vec{F}_{q}(u) \cdot \vec{\nabla} p_{iq} w_{q} 
       -  \sum_{q=1}^{Q_S}p_{jq}\vec{H}_{q}(u^+,u^-)p_{iq}w_{q} ~.
\end{equation}
Whereas the form (\ref{eqn:weak_form_4}) is useful for transport models
when the velocity and conserved fields are easily disentangled, 
equation (\ref{eqn:weak_form_5}) is applicable to
more general flux constructs.

We note that the inverse of the mass matrix appearing in equations 
(\ref{eqn:weak_form_4}) and (\ref{eqn:weak_form_5}) depend only on
the basis functions and Jacobian transformations mapping 
reference elements to actual physical cell geometries. It can thus be computed once
at the start of the simulation and stored to save computational time. Of course it
would have to be recomputed and updated each time the mesh changes by AMR, AOR or
grid motion. But since the matrix elements are entirely local, this can be done on a cell-by-cell
and as needed basis. It does not have to be recomputed globally every cycle across the entire grid.

\subsection{Artificial Viscosity}

We have implemented several variations of an artificial viscosity method for regularizing
shock discontinuities. All versions are conservative in nature and
based on previous work rooted in entropy-based shock detection models \citep{Hartmann02,Guermond11}
but modified here to work with relativistic MHD. The advantage of artificial
viscosity, compared to slope limiting discussed in the next section, is that
it easily generalizes to unstructured grids, to multiple dimensions, to high order
finite elements, and to adaptive mesh and/or order refinement. 
Viscosity is evaluated on each node within a cell, so it is in effect
applied sub-zonally and respects high order compositions of cell elements. Of course
its dissipative nature has to be taken into account when choosing
parameters such as the shock detection threshold, and strength of dissipation.

Artificial viscosity is introduced as a flux conservative, covariant, Laplacian
source term added to the right-hand side of each evolution equation of the form
\begin{equation}
\partial_i(\epsilon_u~\nu(R({\cal U}), \delta J({\cal F}^\pm))~
           \sqrt{-g} g^{ij} \partial_j u) ~,
\end{equation}
where $u$ is the evolved field, $\epsilon_u$ is a constant 
that can differ for each evolved field based on, e.g., (magnetic)
Prandtl number scaling, and $\nu(R({\cal U}), \delta J({\cal F}^\pm))$ is the 
viscosity coefficient that depends on shock jump detection algorithms across cell interfaces
$\delta J({\cal F}^\pm)$ and entropy residuals within zone elements $R({\cal U})$.
The essential viscosity formulation singles out regions of high
entropy production by employing three different detection algorithms to define
reasonable quantitative measures of viscous heating and combines these measures into a
viscosity coefficient that triggers locally over shocks. The three detection functions
are based on: (1) calculating zonal residuals from the transport of an effective entropy function
$R({\cal U}) = \partial_t{\cal U} + \partial_i F^i({\cal U})$; (2) computing flux discontinuities
across cell interfaces $\delta J({\cal F}^\pm)$; and (3) providing an
upper bound determined by the maximum local wave speed in each element, $V_{max}$. The viscosity
is then chosen by
\begin{equation}
\nu(R({\cal U}), \delta J({\cal F}^\pm)) = 
      \text{min}\left( C_l \ell V_{max}, ~\text{max}\left(
                       C_q \ell \frac{\delta J}{||J_N||}, ~C_q \ell^2 \frac{R}{||R_N||}
                       \right) \right) ~.
\end{equation}
$||J_N||$ and $||R_N||$ are locally constructed normalization factors, 
$\ell=\Delta x/p$ is the cell width reduced by the basis order (or equivalently the
distance between nodal sub-grid elements), and
$C_l$ and $C_q$ are the linear and quadratic viscosity
coefficients typically in the ranges $C_l \in [0.1,0.5]$ and $C_q \in [0.2, 1.0]$.
Numerous options for ${\cal U}$ provide reasonable zonal residuals, including
entropy ($\propto P \rho^{1-\Gamma}$), relativistic enthalpy,
stress energy tensor $T^0_0$, and enthalpy scaled mass density. Interface jump detection
is sensed by comparing flux discontinuities in the fields selected for residual
evaluation across cell faces projected into cell face normals $\delta({\cal U} V^i N_i)$.
These zonal residual and interface jump calculations are
typically normalized by the residual fields (e.g., entropy, enthalpy) averaged over zone quadratures,
but can also be normalized by the minimum or maximum quadrature values if the viscosity
needs to be strengthened or weakened.

\subsection{Slope Limiting}

Another option we developed for suppressing spurious oscillations near sharp features is slope limiting. Our 
implementation uses a least squares slope formulation in each cell and applies 
(optional) limiting to either primitive, conserved, or characteristic fields
with a traditional minmod operator.
Specifically in each zone we set up and solve the following least squares problem:
\begin{equation}
  \begin{bmatrix}
   x_0 & y_0 & z_0 \\
   x_1 & y_1 & z_1 \\
   \vdots & \vdots & \vdots \\
   x_n & y_n & z_n
  \end{bmatrix}
    \vec{\delta u} = 
  \begin{bmatrix}
   u_0  \\
   u_1  \\
   \vdots  \\
   u_n 
  \end{bmatrix}	,
\end{equation}
where $(x_i, y_i, z_i)$ are the coordinates of the $i$th support node in the zone,
$u_i$ are the values of the field to be limited at each node, and matrix inversion
gives the least squares solution for the slope vector $\vec{\delta u}$.

Limiting is applied to these slopes along each dimension by 
\begin{equation}
\delta u_{L} = \text{minmod}(\delta u, 
                             \beta (\overline{u} - \overline{u}^-), 
                             \beta (\overline{u}^+ - \overline{u}))	~,
\end{equation}
with
\begin{equation} \label{minmod}
\text{minmod}(a,b,c) = 
  \begin{cases}
      \text{sign}(a) \text{min}(|a|,|b|,|c|) & 
          \text{if}\quad\text{sign}(a) = \text{sign}(b) = \text{sign}(c) \\[0.5em]
      0 & \text{otherwise} ~.
    \end{cases}
\end{equation}
Here $\overline{u}$ denotes the integral of support fields over zone quadratures
on the unit reference element, effectively a quadrature weighted average,
and $\overline{u}^\pm$ denotes the weighted average in neighboring zones
along the positive and negative directions. 
The parameter $\beta \in [0.5,1]$ sets the amount of limiting to be used.

We also implement a bounded version of this limiter in the spirit of
\citet{Cockburn89b} and \citet{Schaal15}, where the central difference
slope is first evaluated against a threshold parameter before applying the minmod operator
\begin{equation} \label{bound_minmod}
\text{minmod}_B(a,b,c) = 
  \begin{cases}
      a & \text{if}\quad |a| \leq M \\
        \text{minmod}(a,b,c) & \text{otherwise}	~,
    \end{cases}
\end{equation}
allowing the user to set a threshold slope below which the limiter will not activate. The 
parameter $M$ depends upon (and is sensitive to) several factors, such as zone size and the 
maximum expected curvature near smooth extrema in the solution. Its optimal
value is in general determined empirically through trial and error.

We note that the combination of the local nature of DG finite elements, the least squares
approach for calculating slopes, and quadrature folding of high order solutions to low
order bases allow these limiting procedures to work easily for any basis order and with
adaptive order refinement. Adaptive mesh refinement too is as easily accommodated
with the additional caveat that if a neighboring zone is on a different
refinement level, $\overline{u}$ is averaged over all of the children in that zone.

\subsection{Time Integration}
\label{subsec:timeint}

The preferred high (greater than second) order time discretization method in 
{\sc Cosmos++} has been a
low-storage version of the forward Euler method \citep{Shu88}.  
In this method, the solution for a generic differential equation,
represented as $\partial_t U = L(U)$, at any stage $i$ can be expressed as
\begin{align}
U^{(i)} = \eta_{i-1} U^{(0)} + (1-\eta_{i-1})[U^{(i-1)} + \Delta t^n L(U^{(i-1)})] ~,
\label{eqn:feuler}
\end{align}
where $U^{(0)} = U^n$ is the solution at $t=t^n$.  
Solutions at any stage $i$ can thus be constructed from the initial solution $U^{(0)}$
and the results of advancing the previous stage, $i-1$.  The coefficients, $\eta_i$,
for the three lowest orders are: $\eta_0=0$ for first order,
$\eta_0=0$, $\eta_1=1/2$ for second, and 
$\eta_0=0$, $\eta_1=3/4$, $\eta_2=1/3$ for third.
Unfortunately, coefficients have not been found to extend the low-storage 
Euler method to higher order.  \citet{Spiteri02} speculate that no such coefficients 
exist. In addition, no 4-stage, 4th order method has been found \citep{Gottlieb98}.

To extend {\sc CosmosDG}
to fourth order we therefore consider an alternative five-stage,
strong-stability-preserving Runge-Kutta (SSPRK) method 
presented in \citet{Spiteri02}.  Generically, an $s$-stage, explicit Runga-Kutta 
method can be written
\begin{align}
U^{(0)} &= U^n , \\
U^{(i)}  &= \sum_{k=0}^{i-1} (\alpha_{ik}U^{(k)} + \Delta t \beta_{ik} L(U^{(k)})) , \hspace{0.1in} \text{for} \; i=1,2,...,s , \label{equation:generic_rk} \\
U^{n+1} &= U^{(s)} .
\end{align}
We only consider cases where the constants $\alpha_{ik} \ge 0$, $\beta_{ik} \ge 0$, 
and $\alpha_{ik} = 0$ if $\beta_{ik} = 0$. For the weighting 
coefficients to be consistent, the $\alpha_{ik}$ must satisfy $\sum_{k=0}^{i-1} \alpha_{ik} = 1$.  
This Runga-Kutta method is strong stability preserving provided
\begin{equation}
\Delta t \le c \Delta t_\mathrm{FE},
\label{eq:timestep}
\end{equation}
where
\begin{equation}
c \equiv \underset{i,k}{\text{min}} \frac{\alpha_{ik}}{\beta_{ik}}  ~,
\end{equation}
and $\Delta t_\mathrm{FE}$ comes from stability requirements on the forward Euler timestep.
For DG methods this is calculated as $\Delta t_\mathrm{FE} = {c_{FL}}/(p+1)$ 
times the minimum estimated stability timestep over all physics packages,
where $p$ is the basis order. The
Courant constant ${c_{FL}}$ is typically set to 0.5 or less.
The coefficients $\alpha_{ik}$ and $\beta_{ik}$ are displayed in Table \ref{table:rk_constants_five_stage},
using standard (row, column) indexing, along with the
corresponding timestep coefficients, $c$, for the five-stage method at convergence 
orders 2, 3, and 4.  It is possible to write coefficients for a first-order, five-stage scheme, but
since it offers no efficiency advantage over a standard Euler scheme, it is not implemented.

\begin{table}
  \centering
  \caption{Coefficients for Five-Stage, Strong-Stability-Preserving Runge-Kutta Integration Method
  \label{table:rk_constants_five_stage}}
  \begin{tabular}{|c|c|c|ccccc|}
    \tableline
    Order       & $c$               & \multicolumn{6}{c|}{}                                                                                                        \\
    \tableline
    2          & 4                &               & 1                &                  &                  &                  &                  \\
                &                   &               & 0                & 1                &                  &                  &                  \\
                &                   &  $\alpha_{ik}$ & 0                & 0                & 1                &                  &                  \\
                &                   &               & 0                & 0                & 0                & 1                &                  \\
                &                   &               & 0.2              & 0                & 0                & 0                & 0.8              \\
    \cline{3-8}
                &                   &               & 0.25             &                  &                  &                  &                  \\
                &                   &               & 0                & 0.25             &                  &                  &                  \\
                &                   & $\beta_{ik}$  & 0                & 0                & 0.25             &                  &                  \\
                &                   &               & 0                & 0                & 0                & 0.25             &                  \\
                &                   &               & 0                & 0                & 0                & 0                & 0.2              \\
    \tableline
    3           & 2.65062919294483  &  & 1                &                  &                  &                  &                  \\
                &                   &               & 0                & 1                &                  &                  &                  \\
                &                   & $\alpha_{ik}$ & 0.56656131914033 & 0                & 0.43343868085967 &                  &                  \\
                &                   &               & 0.09299483444413 & 0.00002090369620 & 0                & 0.90698426185967 &                  \\
                &                   &               & 0.00736132260920 & 0.20127980325145 & 0.00182955389682 & 0                & 0.78952932024253 \\
    \cline{3-8}
                &                   &   & 0.37726891511710 &                  &                  &                  &                  \\
                &                   &               & 0                & 0.37726891511710 &                  &                  &                  \\
                &                   & $\beta_{ik}$ & 0                & 0                & 0.16352294089771 &                  &                  \\
                &                   &               & 0.00071997378654 & 0                & 0                & 0.34217696850008 &                  \\
                &                   &               & 0.00277719819460 & 0.00001567934613 & 0                & 0                & 0.29786487010104 \\
    \tableline
    4           & 1.50818004975927  &  & 1                &                  &                  &                  &                  \\
                &                   &               & 0.44437049406734 & 0.55562950593266 &                  &                  &                  \\
                &                   &  $\alpha_{ik}$ & 0.62010185138540 & 0                & 0.37989814861460 &                  &                  \\
                &                   &               & 0.17807995410773 & 0                & 0                & 0.82192004589227 &                  \\
                &                   &               & 0.00683325884039 & 0                & 0.51723167208978 & 0.12759831133288 & 0.34833675773694 \\
    \cline{3-8}
                &                   &   & 0.39175222700392 &                  &                  &                  &                  \\
                &                   &               & 0                & 0.36841059262959 &                  &                  &                  \\
                &                   & $\beta_{ik}$ & 0                & 0                & 0.25189177424738 &                  &                  \\
                &                   &               & 0                & 0                & 0                & 0.54497475021237 &                  \\
                &                   &               & 0                & 0                & 0                & 0.08460416338212 & 0.22600748319395 \\
    \tableline
  \end{tabular}
\end{table}

As we mentioned, the five-stage method is required to achieve fourth-order convergence.  
However, \citet{Spiteri02} have shown that the five-stage scheme can have advantages at 
lower order, too.  This is because the effective timestep that results from 
equation (\ref{eq:timestep}) can be considerably larger than $\Delta t_\mathrm{FE}$.  
So, although the five-stage method is undoubtably more expensive per full update 
cycle, it can require far fewer total cycles.  As an example, a simple two-stage, 
second-order scheme will be able to step forward $\Delta t_\mathrm{FE}$ each cycle, 
whereas a five-stage, second-order scheme will be able to step forward $4 \Delta t_\mathrm{FE}$.  
Thus, although the five-stage scheme requires $5/2$ more work per cycle, it goes 4 times 
further each cycle, making it $[(4/1)/(5/2) - 1]\times 100\% = 60\%$ more efficient over the full evolution.

It is worth mentioning that in order for any SSPRK scheme to be 
implemented as a low-storage method, the constants $\alpha_{ik}$ and $\beta_{ik}$ 
must be such that no intermediate stage solutions are required in the final stage. 
A low-storage scheme would require $\alpha_{ik}=0$ for $k < i-1$ whenever $i<s$ 
and for $k=1, ..., s-1$ for $i = s$.  Similarly, it would 
require $\beta_{ik}=0$ for $k < i-1$ for any $i$.  We see that, of the five-stage 
options in Table \ref{table:rk_constants_five_stage}, only the second-order 
one could be done using the low-storage approach.

\section{Code Tests}

Some of the code tests presented here are taken from our earlier papers which first 
introduced {\sc Cosmos++}.  If a direct comparison of our latest results to analytic or
published numerical solutions is not made explicit in the following sections, the reader 
is referred to \citet{Anninos03a} and \citet{Anninos05} to establish comparisons 
between DG and two different flavors of FV solutions:
high resolution shock capturing (HRSC), and non-oscillatory
central difference (NOCD).

\subsection{Linear Waves}
\label{sec:pert}

We begin by performing convergence studies of smooth linearized perturbation waves, reproducing
a subset of special relativistic tests proposed by \cite{Sadowski14} (see also \cite{Fragile14}). In particular
we consider the first three cases of their Table 2 corresponding to sonic,
fast magnetosonic, and slow magnetosonic waves. 
These solutions are represented by the real parts of the eigenmodes
\begin{equation}
q^a = \text{Re}\left( q_0^a + \delta q^a e^{i(\omega t - kx)} \right) 	~,
\label{eqn:pert}
\end{equation}
for each fluid variable and magnetic field component represented by superscript $a$.
The subscript $0$ denotes the unperturbed background values, $\delta q^a$ are the
perturbation eigenvectors, and $\omega$ and $k$ are the complex frequency and
wave number respectively. The unperturbed background values are the same
in all three tests: $\rho_0=1$, $u_0^x=u_0^y=0$, $B_0^x=B_0^y=0.100758544372$.
The sound speed in the background gas is $c_{s,0}=0.1$ and for a $\Gamma=5/3$ ideal gas
this gives an internal energy density of 
$e_0=\rho_0\epsilon_0=9.13705584\times10^{-3}$.
When magnetic fields are present, the background field is
evenly split between $x$ and $y$ components such that the Alfv\'en speed is $v_{A,0}=0.2$.
The first order perturbation constants $\delta q^a$ are provided in Table \ref{tab:pert_analytic}.
The wave number is taken as $k=2\pi/L$, where $L$ is the grid length set to unity
for all tests. All calculations are run to $t=6\pi/\text{Re}(\omega)$ corresponding to 
three complete wave periods, enforcing periodicity on all fields along external boundary zones.
A third order, forward Euler time integration is used for the evolutions.

\begin{figure}
\includegraphics[width=0.6\textwidth]{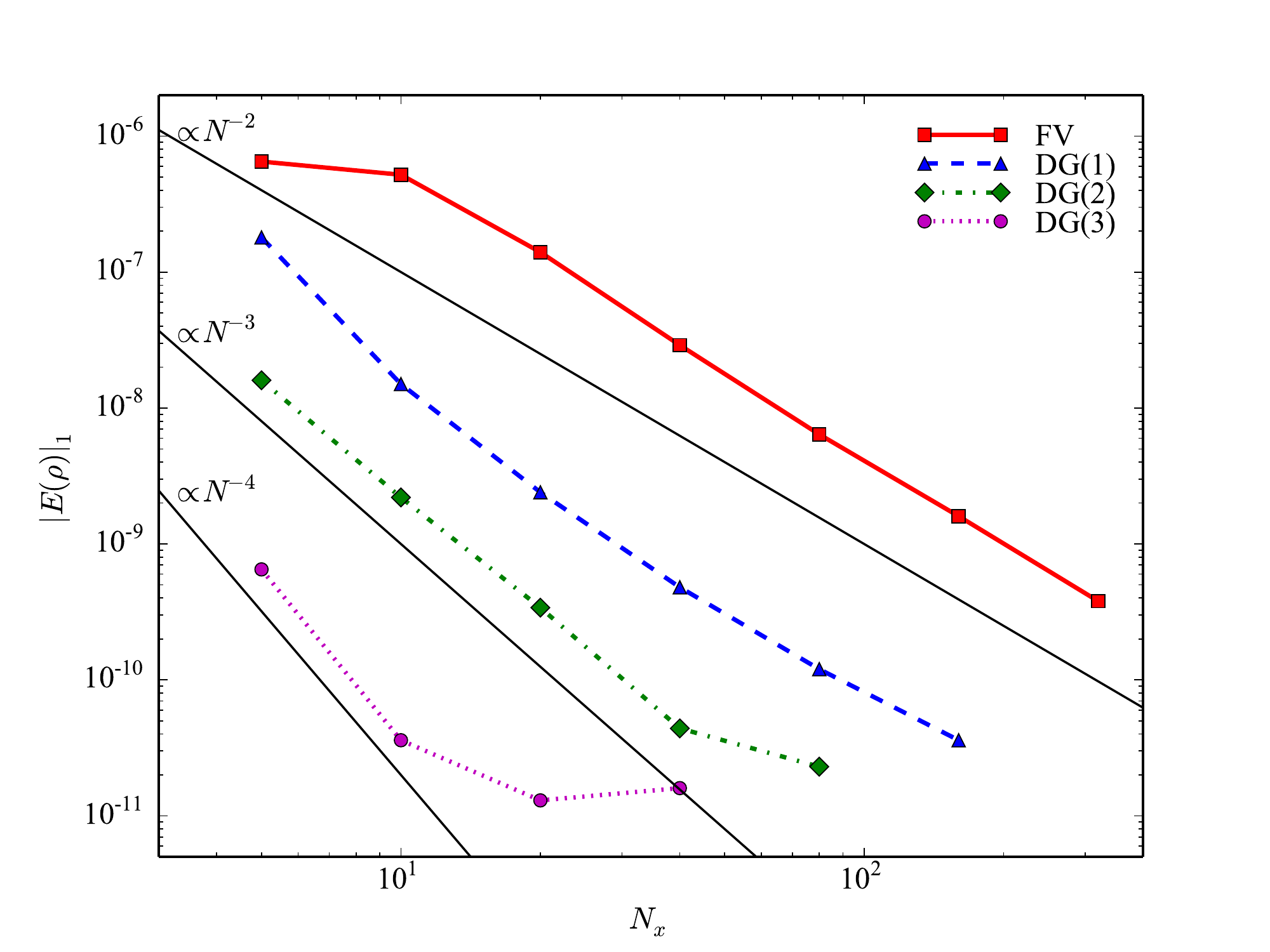}
\caption{Plot of the L1-norm errors in mass density for the slow magnetosonic wave perturbation test
in Table \ref{tab:pert_error}.}
\label{fig:pert}
\end{figure}

L1-norm errors of the mass density are displayed in Table
\ref{tab:pert_error} as a function of $N_x$, the number of zones along
the propagation $x$-direction, comparing results of DG(1), DG(2) and DG(3) (2nd, 3rd, and 4th
order) solutions against the finite volume (2nd order) version of {\sc Cosmos++}.
Notice DG(1) and FV results converge like second order as expected,
but interestingly the DG solutions are about 20 to 50 times more accurate overall. Additionally,
the DG(2) and DG(3) results also converge at their expected rates, exhibiting
eight and sixteen fold increases in accuracy with each doubling of zones.
We point out that convergence is evaluated against first order perturbation solutions,
and are thus valid only to second order contributions,
$\delta^{(2)}\rho/\rho \approx \text{few} \times10^{-11}$.
This accounts for the flattening of convergence curves in Table \ref{tab:pert_error}, 
also shown graphically in Figure \ref{fig:pert}
for the slow magnetosonic case. The fourth order DG(3) method very quickly
reaches this level of accuracy and the convergence curve
saturates after just ten zones. By contrast, more
than 1300 zones would be required to achieve this level of accuracy with traditional
second order methods.

\begin{table}
  \centering
  \caption{Eigenmode Solutions of Linear Sonic and Magnetosonic Waves
  \label{tab:pert_analytic}}
  \begin{tabular}{c|cccc}
    \tableline
                     & \text{sonic}                 & \text{fast ~MHD}             & \text{slow ~MHD} \\
    \tableline
    $\delta\rho$     & $10^{-6} + 0i$               & $10^{-6} + 0i$               & $10^{-6} + 0i$   \\
    $\delta e  $     & $1.52284\times10^{-8} + 0i$  & $1.52284\times10^{-8} + 0i$  & $1.52284\times10^{-8} + 0i$  \\
    $\delta u^x$     & $10^{-7} + 0i$               & $1.60294\times10^{-7} + 0i$  & $6.17707\times10^{-8} + 0i$  \\
    $\delta u^y$     & $0 + 0i$                     & $-9.79087\times10^{-8}+ 0i$  & $1.00118\times10^{-7} + 0i$  \\
    $\delta B^y$     & $0 + 0i$                     & $1.62303\times10^{-7}+ 0 i$  & $-6.25516\times10^{-8} + 0i$ \\
    $\omega    $     & $0.628319 + 0i$              & $1.00716 + 0 i$              & $0.388117 + 0i$ \\
    \tableline
  \end{tabular}
\end{table}

\begin{table}
  \centering
  \caption{L$1$-norm errors of mass density for Linear Wave tests
  \label{tab:pert_error}}
  \begin{tabular}{l|ccccccc}
    \tableline
                 & $N_x=5$             & $N_x=10$            & $N_x=20$             
                 & $N_x=40$            & $N_x=80$            & $N_x=160$            & $N_x=320$  \\
    \tableline
    FV-sonic     & $6.3\times10^{- 7}$ & $3.5\times10^{- 7}$ &  $1.0\times10^{- 7}$ 
                 & $2.5\times10^{- 8}$ & $6.2\times10^{- 9}$ &  $1.5\times10^{- 9}$ &  $3.8\times10^{-10}$ \\
    FV-fast      & $6.4\times10^{- 7}$ & $3.6\times10^{- 7}$ &  $1.0\times10^{- 7}$ 
                 & $2.5\times10^{- 8}$ & $6.2\times10^{- 9}$ &  $1.5\times10^{- 9}$ &  $3.5\times10^{-10}$ \\
    FV-slow      & $6.5\times10^{- 7}$ & $5.2\times10^{- 7}$ &  $1.4\times10^{- 7}$ 
                 & $2.9\times10^{- 8}$ & $6.4\times10^{- 9}$ &  $1.6\times10^{- 9}$ &  $3.8\times10^{-10}$ \\
    \tableline
    DG(1)-sonic  & $2.2\times10^{- 7}$ & $4.2\times10^{- 8}$ &  $5.8\times10^{- 9}$
                 & $1.3\times10^{- 9}$ & $3.3\times10^{-10}$ &  $8.3\times10^{-11}$ &   \\
    DG(1)-fast   & $2.1\times10^{- 7}$ & $3.9\times10^{- 8}$ &  $5.5\times10^{- 9}$
                 & $1.2\times10^{- 9}$ & $3.1\times10^{-10}$ &  $8.4\times10^{-11}$ &   \\
    DG(1)-slow   & $1.8\times10^{- 7}$ & $1.5\times10^{- 8}$ &  $2.4\times10^{- 9}$
                 & $4.8\times10^{-10}$ & $1.2\times10^{-10}$ &  $3.6\times10^{-11}$ &   \\
    \tableline
    DG(2)-sonic  & $1.2\times10^{- 8}$ & $1.3\times10^{- 9}$  &  $1.6\times10^{-10}$ 
                 & $2.1\times10^{-11}$ & $1.1\times10^{-11}$ &   &   \\
    DG(2)-fast   & $1.2\times10^{- 8}$ & $1.3\times10^{- 9}$  &  $1.6\times10^{-10}$ 
                 & $2.8\times10^{-11}$ & $3.1\times10^{-11}$ &   &   \\
    DG(2)-slow   & $1.6\times10^{- 8}$ & $2.2\times10^{- 9}$  &  $3.4\times10^{-10}$ 
                 & $4.4\times10^{-11}$ & $2.3\times10^{-11}$ &   &   \\
    \tableline
    DG(3)-sonic  & $1.5\times10^{- 9}$ & $1.4\times10^{-10}$  &  $2.0\times10^{-11}$
                 & $1.1\times10^{-11}$ &  &   &   \\
    DG(3)-fast   & $1.5\times10^{- 9}$ & $1.3\times10^{-10}$  &  $3.5\times10^{-11}$
                 & $3.3\times10^{-11}$ &  &   &   \\
    DG(3)-slow   & $6.5\times10^{-10}$ & $3.6\times10^{-11}$  &  $1.3\times10^{-11}$
                 & $1.6\times10^{-11}$ &  &   &   \\
    \tableline
  \end{tabular}
\end{table}

\subsection{Alfv\'en Shearing Modes}

\citet{DeVilliers03} described a class of linear Alfv\'en wave solutions
that test magnetic fields subject to transverse or shearing
mode perturbations. Under the conditions of small amplitude perturbations with
a fixed background magnetic field $B^x$ and constant velocity $V^x$ in Minkowski
spacetime, the transverse velocity and field components can be written
\begin{equation}
V^y = \frac{1-\zeta\chi}{2} f(x-v_A^- t) + \frac{1+\zeta\chi}{2} f(x - v_A^+ t) ~, 
\end{equation}
\begin{equation}
B^y = \frac{\zeta}{2}(f(x-v_A^- t) - f(x-v_A^+ t)) ~,
\end{equation}
with
\begin{equation}
\zeta = \frac{B^x(1+\eta^2)}{\eta\sqrt{\eta^2+W^{-2}}}	~, \qquad 
\eta^2=\frac{|B|^2}{\rho h W^2} ~,
\end{equation}
\begin{equation}
\chi = -\frac{\eta^2 V^x}{B^x(1+\eta^2)}  	~, \qquad 
\beta = \frac{2 P}{|B|^2} ~,
\end{equation}
and Alfv\'en speeds
\begin{equation}
v_A^{\pm} = \frac{V^x \pm \eta \sqrt{\eta^2+W^{-2}}}{1+\eta^2} ~.
\end{equation}

We consider two cases: a stationary
background ($V^x=0$) where pulses travel in opposite directions with equal
amplitudes (case ALF-1), and a moving background ($V^x = 0.1 c$) where pulses split into
asymmetrical waves (case ALF-2). These cases correspond to models ALF1 and ALF3
of \citet{DeVilliers03}.
The fluid is initialized with uniform unit density,
zero transverse magnetic field components $B^y=B^z=0$,
specific energy $\epsilon = 10^{-2}$, and ideal gas constant
$\Gamma=5/3$. The longitudinal field component $B^x$ is set by
the parameter $\beta$: 0.001 for ALF-1,
0.01 for ALF-2. The transverse velocity
function $f(x,t)$ is initialized as a square pulse: $V^y=f(x,0)=10^{-3}c$ 
for $1<x<1.5$, $V^y=f(x,0)=-10^{-3}c$ for $1.5 \le x < 2$, and zero everywhere else
(the grid length runs from 0 to 3 units, with periodic boundary conditions). The Alfv\'en wave speeds 
are $|v_A^\pm|=0.96c$ for ALF-1, and
$v_A^+ = 0.79c$ and $v_A^- = -0.71c$ for ALF-2.

\begin{figure}
\includegraphics[width=0.48\textwidth]{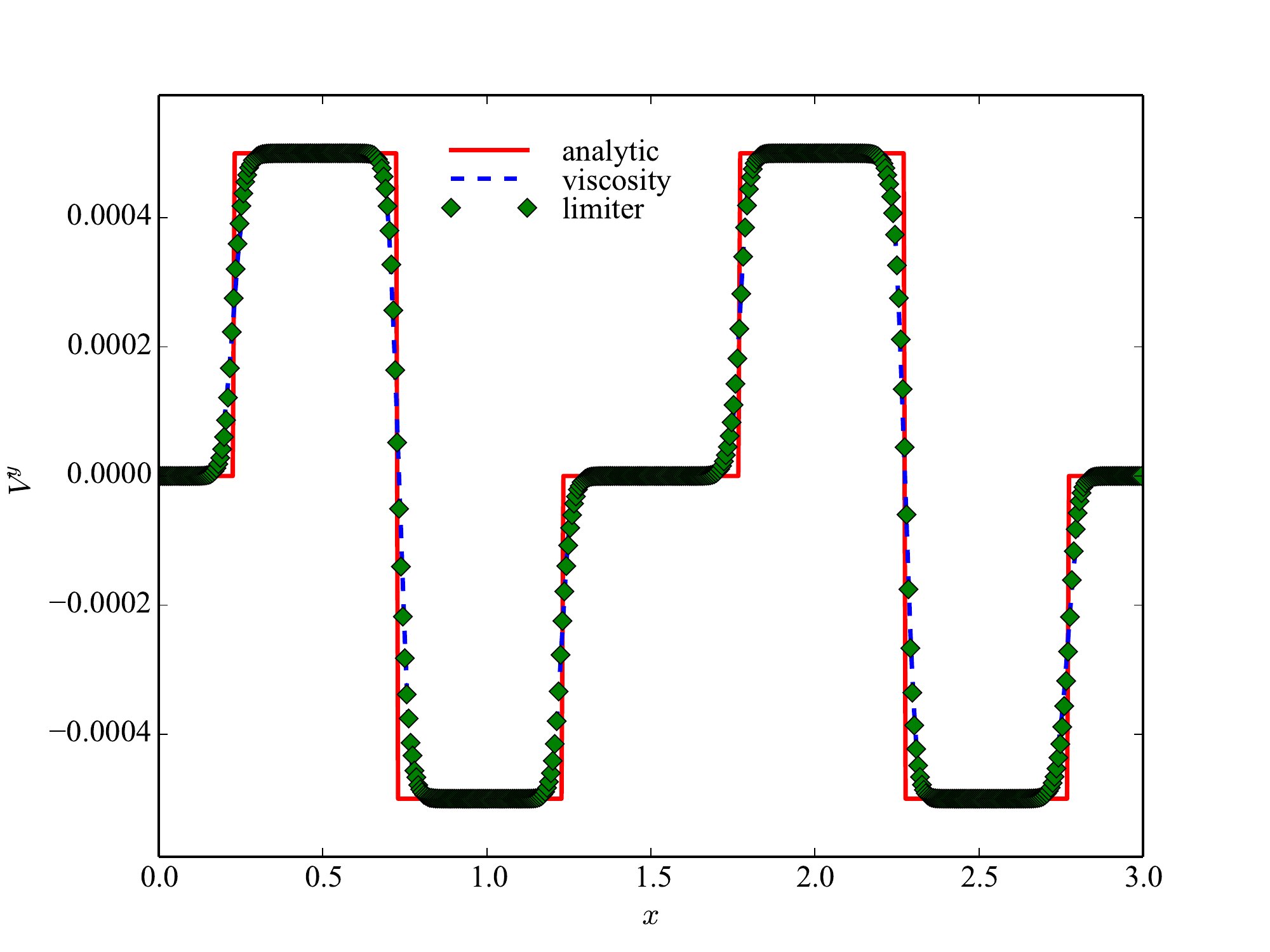}
\includegraphics[width=0.48\textwidth]{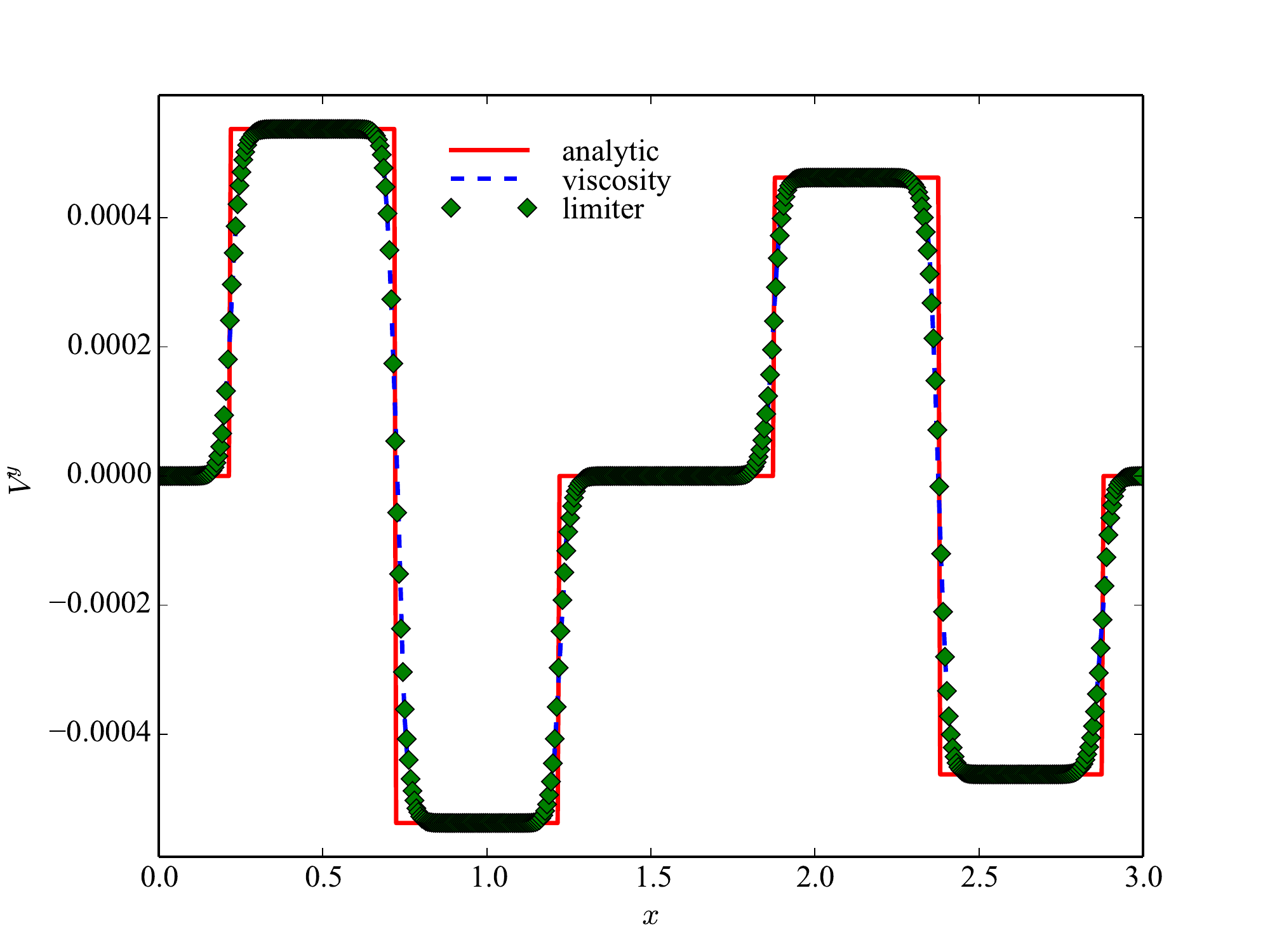}
\caption{Transverse velocity $V^y$ for the ALF-1 (left) and ALF-2 (right) shearing mode tests
with 512 zone resolution.
Solid lines are the analytic solutions given in the text, 
dashed lines are the DG(1) solutions using entropy viscosity for
discontinuity capturing, and the diamonds are the
DG(1) solutions with the slope limiter.}
\label{fig:shear_caseAnB}
\end{figure}

Numerical results are plotted in 
Figure \ref{fig:shear_caseAnB},
where we compare analytic to DG(1) solutions.
Two calculations are shown for both cases: one using entropy
viscosity to capture the discontinuities, the second using the slope limiter.
Solutions for the two discontinuity capturing approaches are very similar,
and they both match the finite volume calculations.
All solver permutations (DG, FV, viscosity, limiter) reproduce the plateau values
to better than 0.002\%, and converge globally to the analytic solution
at rates close to unity.

\begin{figure}
\includegraphics[width=0.6\textwidth]{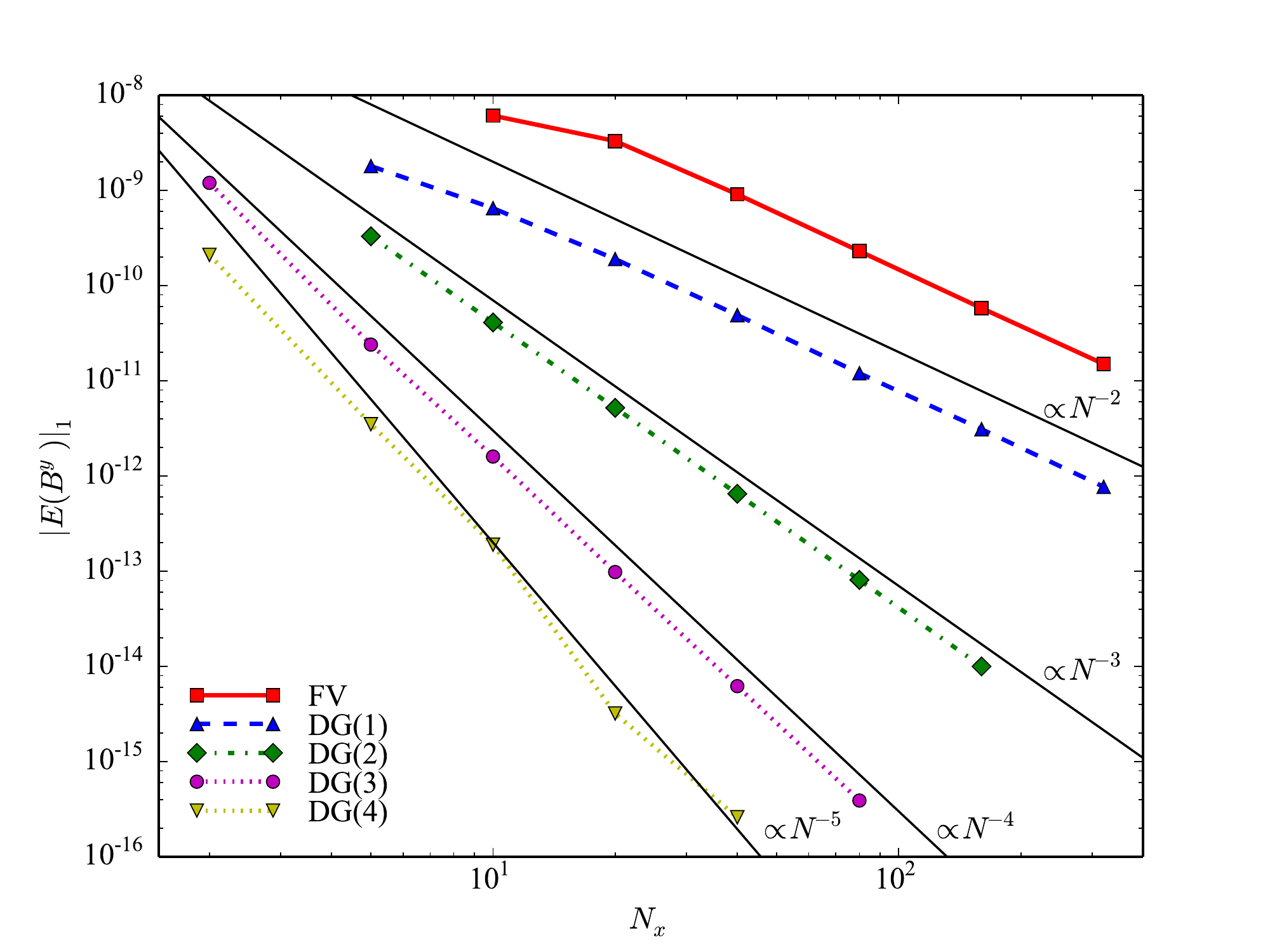}
\caption{Plot of the L1-norm errors in $B^y$ for the smooth Alfv\'en shearing mode perturbation test.}
\label{fig:Alfven_error}
\end{figure}

Taking advantage of the semi-analytic nature of this solution, we 
additionally consider a smooth wave form for the function
$f(x-v_A^{\pm}t) = f_0\sin(2\pi(x - v_A^{\pm}t)/L)$,
with small amplitude $f_0 = 10^{-8}$ to expand the perturbation regime. 
This allows us to perform convergence studies similar to those
conducted in section \ref{sec:pert}.
Although this problem is not as rigorous a test for hydrodynamics as those presented 
above, it is nonetheless a useful diagnostic of magnetically dominated flows.
For these series of tests we use
the 4th order, five-stage Runge-Kutta integrator and extend our testing to
include DG(4), 5th order spatial discretization. L1-norm errors for $B^y$ are presented
in Table \ref{tab:Alfven_error} and plotted in Figure \ref{fig:Alfven_error}.
As before, all calculations were run for three complete wave periods of the 
fast Alfv\'en mode.
We find convergence rates generally consistent with the spatial order of each scheme:
2nd order for the FV and DG(1) schemes, and order $p+1$ for the DG($p$) methods.

\begin{table}
  \centering
  \caption{L$1$-norm errors in $B^y$ for the smooth Alfv\'en shearing mode test
  \label{tab:Alfven_error}}
  \begin{tabular}{l|ccccccc}
    \tableline
            & $N_x=5$             & $N_x=10$            & $N_x=20$             
            & $N_x=40$            & $N_x=80$            & $N_x=160$           & $N_x=320$  \\
    \tableline
    FV      & $1.0\times10^{- 8}$ & $6.1\times10^{- 9}$ & $3.3\times10^{- 9}$ 
            & $9.1\times10^{-10}$ & $2.3\times10^{-10}$ & $5.8\times10^{-11}$ & $1.5\times10^{-11}$ \\
    DG(1)   & $1.8\times10^{- 9}$ & $6.5\times10^{-10}$ & $1.9\times10^{-10}$
            & $4.9\times10^{-11}$ & $1.2\times10^{-11}$ & $3.1\times10^{-12}$ & $7.7\times10^{-13}$ \\
    DG(2)   & $3.3\times10^{-10}$ & $4.1\times10^{-11}$ & $5.2\times10^{-12}$ 
            & $6.5\times10^{-13}$ & $8.1\times10^{-14}$ & $1.0\times10^{-14}$ &   \\
    DG(3)   & $2.4\times10^{-11}$ & $1.6\times10^{-12}$ & $9.8\times10^{-14}$
            & $6.2\times10^{-15}$ & $3.9\times10^{-16}$ &  &   \\
    DG(4)   & $3.5\times10^{-12}$ & $1.9\times10^{-13}$ & $3.2\times10^{-15}$
            & $2.6\times10^{-16}$ &  &   &   \\
    \tableline
  \end{tabular}
\end{table}

\subsection{Hydrodynamic Shocks}

We consider two special relativistic hydrodynamic shock tube tests: 
a relatively mild boost case (HDST-1) with $V = 0.7c$,
and a second higher boost case (HDST-2) with $V = 0.9c$.
These tests set up two different fluid states separated
by a membrane in the middle of the domain that is removed at $t=0$. 
The fluid subsequently evolves to
form a leftward propagating rarefaction wave, and rightward propagating
contact discontinuity and shock wave.
The initial data for HDST-1
is specified as $\rho_L=10$, $p_L=10$, $V^x_L=0$,
to the left of the partition, $\rho_R=1$, $p_R=10^{-2}$, $V^x_R=0$
to the right, and a $\Gamma=4/3$ ideal gas equation of state. 
HDST-2 is similar but with a significantly greater
pressure to the left of the membrane, $p_L=170$, which makes the leading Lorentz
contracted density discontinuity much harder to resolve
on meshes with limited cell resources.

\begin{figure}
\includegraphics[width=0.48\textwidth]{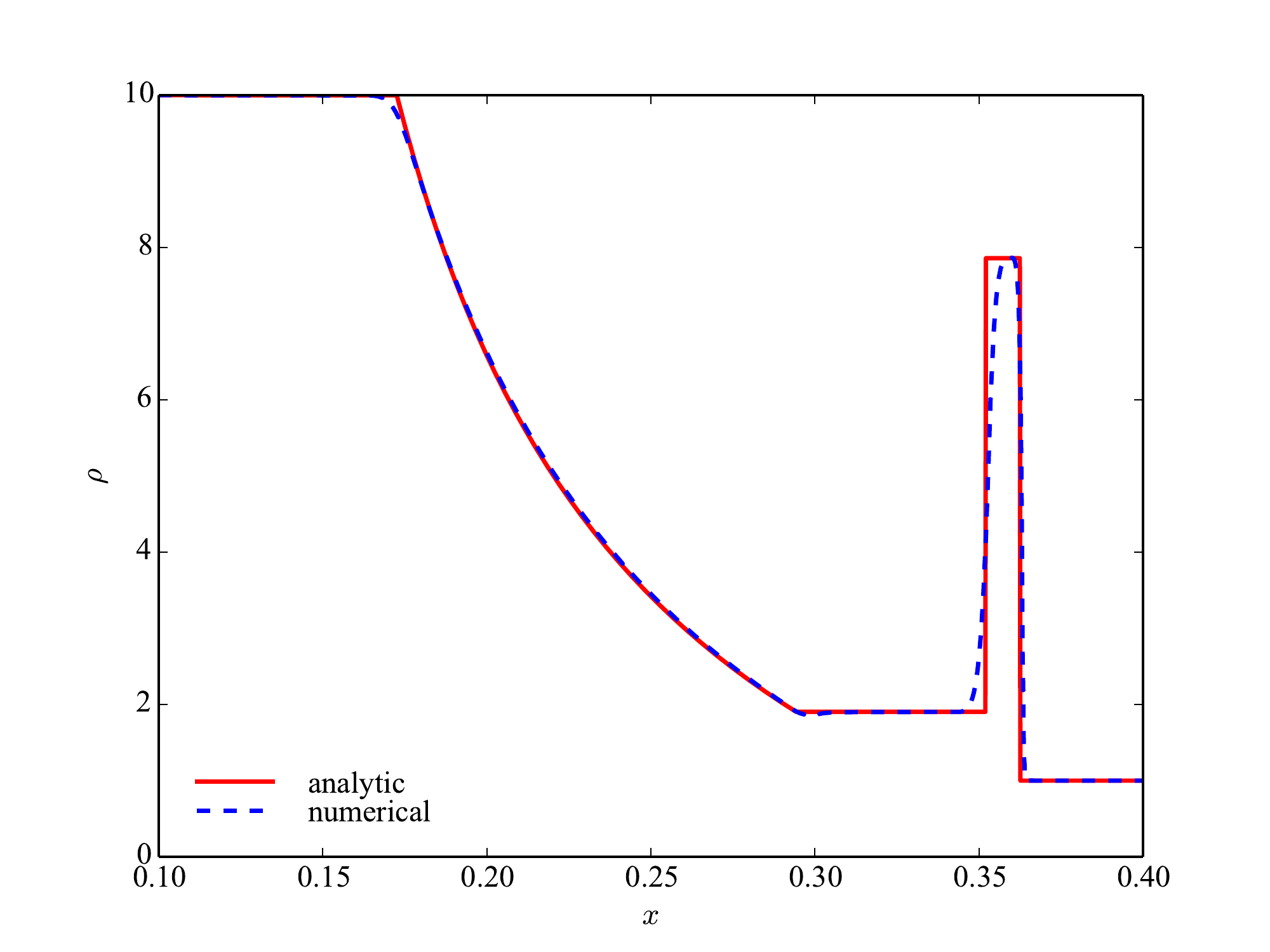}
\includegraphics[width=0.48\textwidth]{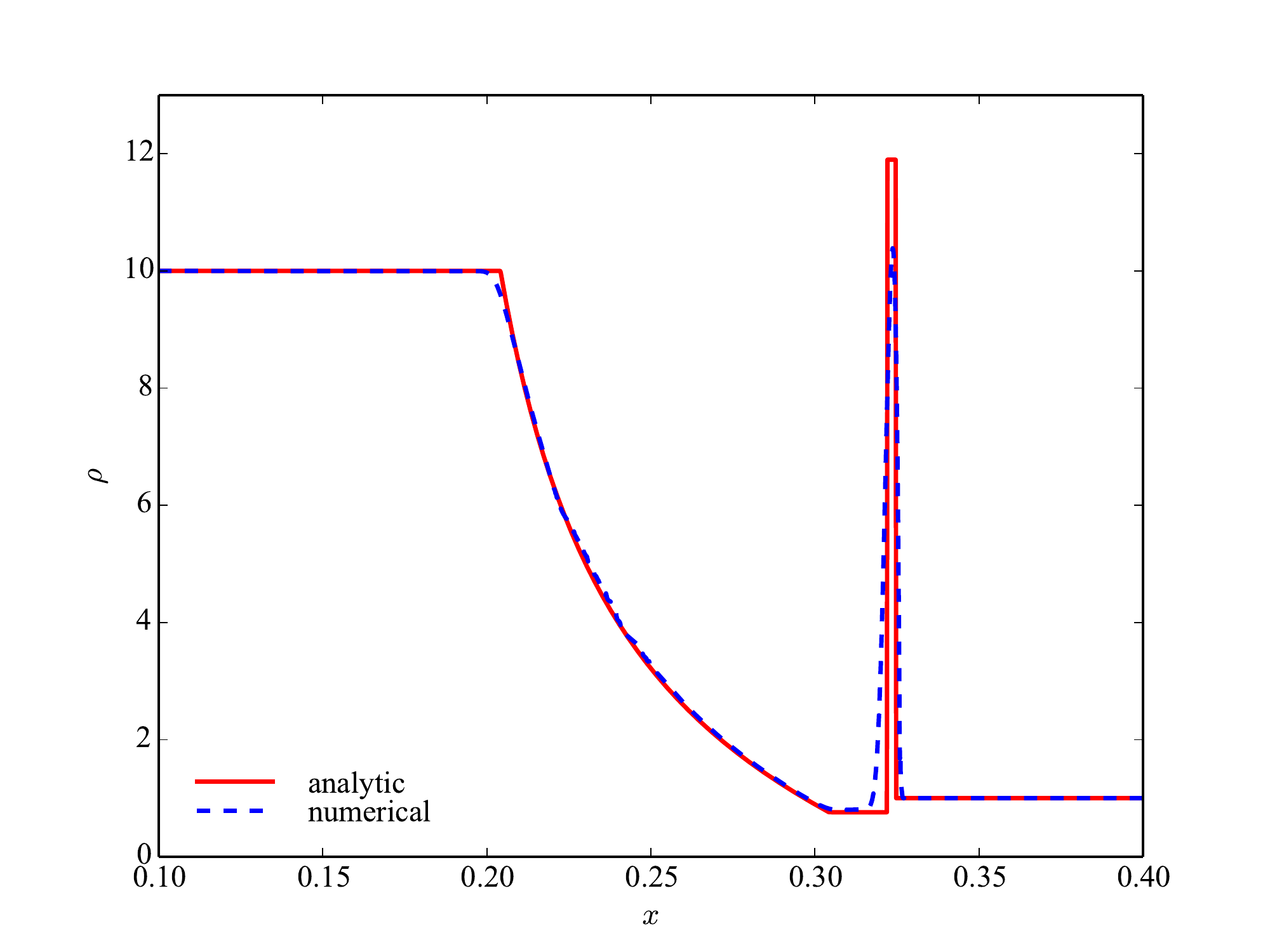}
\caption{Mass density $\rho$ for the hydrodynamic shock tube test HDST-1 (left) at time $t=0.15$,
and HDST-2 (right) at time $t=0.08$.
Solid lines are the analytic Riemann solutions and dashed lines
are numerical solutions using slope limiting for shock regularization 
in the case of HDST-1, and artificial viscosity in the case of HDST-2.
Numerical solutions are calculated on a domain from 0 to 0.5 with 1280 zones.}
\label{fig:hydroshocks}
\end{figure}

All tests are performed separately with artificial viscosity or slope limiting, imposing flat
(zero gradient) boundary conditions,
and third or fourth order elements (DG(2), DG(3)) to demonstrate the robustness of
the different shock regularization techniques and the application of high order finite elements
to shock problems. L1-norm errors of mass density are shown in Table \ref{tab:hydroshock} for
both cases, both regularizations, and across a range of grid resolutions
to compute convergence rates. Errors are evaluated at a final time of $t=0.06$.
For the viscosity runs, common values of 0.2 and 0.6 are used for the linear and quadratic
coefficients respectively. The slope limiter calculations use a steepness parameter of unity.
Notice that the errors quoted in Table \ref{tab:hydroshock} converge to the analytic solutions
at roughly first order.  This is what is expected for simulations that include strong discontinuities, such as shocks. The corresponding solutions are shown in
Figure \ref{fig:hydroshocks}
where we plot mass densities at $t=0.15$ for HDST-1 and $t=0.08$ for HDST-2.
Solid lines represent the analytic solutions derived by solving the
exact Riemann problem, and dashed lines are the numerical solutions
on grids resolving a domain from zero to 0.5 with 1280 zones. 
The numerical solutions are calculated using slope limiting in the case of
HDST-1 and artificial viscosity in the case of HDST-2.
In the more difficult HDST-2 test, the analytic and calculated shock jump states agree
to about ten percent at the 1280 zone resolution used in producing
Figure \ref{fig:hydroshocks}, but converge linearly with resolution.

\begin{table}
  \centering
  \caption{L$1$-norm errors in mass density for the hydrodynamic shock tube tests
  \label{tab:hydroshock}}
  \begin{tabular}{l|ccccc}
    \tableline
                      & $N_x=320$ & $N_x=640$ & $N_x=1280$ & $N_x=2560$  \\
    \tableline
    HDST-1: limiter   & $0.112$   & $0.063$   & $0.038$    & $0.020$ \\
    HDST-1: viscosity & $0.105$   & $0.065$   & $0.043$    & $0.028$ \\
    \tableline
    HDST-2: limiter   & $0.176$   & $0.105$   & $0.070$    & $0.038$ \\
    HDST-2: viscosity & $0.148$   & $0.094$   & $0.053$    & $0.027$ \\
    \tableline
  \end{tabular}
\end{table}

\subsection{Boosted Shock Collision}

\cite{Anninos05} derived an exact solution for the collision of two boosted
fluids that tests the Lorentz invariance of the code under rigorous non-symmetric conditions,
multiple jump discontinuities, and highly relativistic shocks. 
In the center-of-momentum frame this problem consists of two colliding fluids, one flowing from
the left, one from the right. 
The pre- and post-shock states of the two fluids are defined in the center of mass
(primed) frame by zero post-shock velocities
and pressure equilibrium assuming an infinite strength (cold fluid) approximation:
\begin{equation}
P_\mathrm{post} = \rho_\mathrm{post}(\Gamma-1)(W'_\mathrm{pre} - 1) ~,
\end{equation}
\begin{equation}
\rho_\mathrm{post} = \rho_\mathrm{pre}\frac{1 + \Gamma W'_\mathrm{pre}}{\Gamma-1} ~.
\end{equation}
The observer is then boosted to the right at a specified
velocity so that the shocked region appears to move to the left at very high
velocities, even as they move apart (in opposite directions) in the center-of-momentum frame.
The velocity of the center of mass frame and contact discontinuity is calculated by solving
the nonlinear boost transformation equations.
We do not repeat the derivation
here, but refer the reader to Section 4.1.3 of their paper for a detailed discussion of
the initialization, and their Table 2 where the solutions of three specific cases are recorded. 

\begin{figure}
\includegraphics[width=0.6\textwidth]{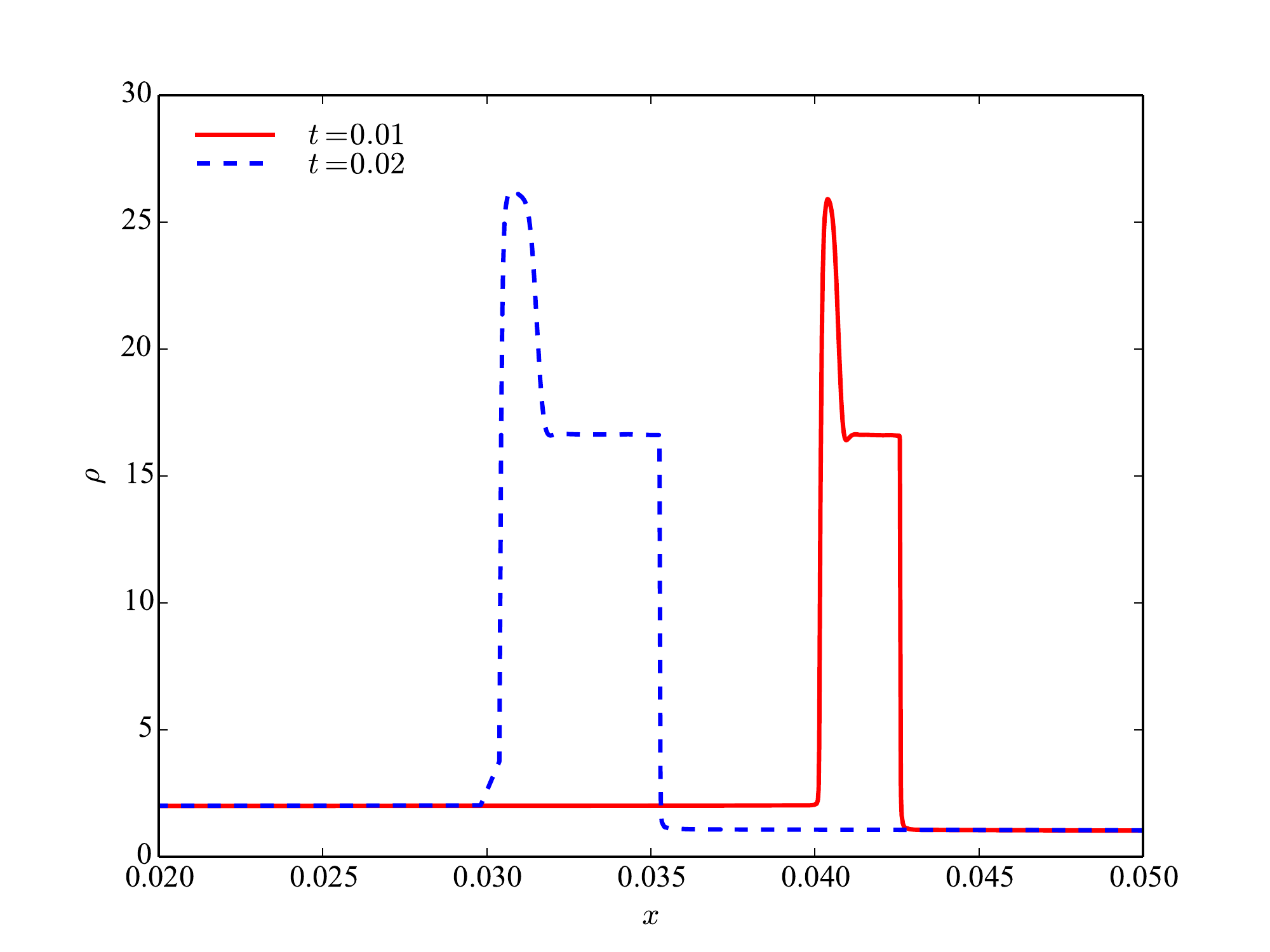}
\caption{Mass density $\rho$ for the ultra-relativistic colliding shock test at two
different times $t=0.01$ (solid) and 0.02 (dashed), showing the fluids moving
to the left at velocity 0.999c.}
\label{fig:collision}
\end{figure}

These tests produce highly relativistic shocks that require extremely fine zoning
to properly capture the jump conditions. We achieve this with adaptive mesh refinement, using up
to 7 levels of refinement on top of a base grid of length 0.06 covered by 80 zones.
In addition we have run these problems with both AMR and AOR in combination
to test the simultaneous use of both refinement techniques, although in practice
high order polynomials are ineffective in these shock dominated cases producing results
essentially identical to low order solutions.
We have reproduced comparable quality solutions for all the cases derived in
\cite{Anninos05}, but show 
representative results in Figure \ref{fig:collision} for the case
where the colliding fluids have different densities.
The two fluids have initial proper densities 
$\rho_1=2$ (left), $\rho_2=1$ (right), pressures $P=10^{-6}$, and adiabatic index $\Gamma=5/3$.
In the center of mass frame, the fluids move in opposite directions, each with $W'=5$. 
The observer is boosted to the right at $W=3$, so the fluid moves at
speeds up to $0.999c$. The two curves in Figure \ref{fig:collision} are 
solutions for mass densities
calculated with artificial viscosity shock capturing and flat boundary conditions at two different times: 
$t=0.01$ (dashed), and $t=0.02$ (solid), showing
the shocked fluid moving to the left. The corresponding solutions with slope limiting 
appear very similar.
For the artificial viscosity solution we find fractional 
density errors of $\approx 5\times10^{-2}$ and $2\times10^{-2}$ in
the higher and lower density post-shock plateaus respectively, and energy density
errors at the contact discontinuity of about $\delta e/e \approx 7\times10^{-2}$.
Errors are slightly better for the slope limiter calculations:
$\delta\rho/\rho \approx 3\times10^{-2}$ ($2\times10^{-3}$) for the high (low)
density plateaus, and $\delta e/e \approx 10^{-3}$ for the post-shock energy density.

\subsection{MHD Shocks}

Next we consider three magneto-hydrodynamic shock tube tests:
the first two are taken from \citet{Komissarov99}, the third is a
relativistic version of the Brio-Wu shock tube \citep{DeVilliers03}. All are initialized with
zero velocities, and discontinuities separating the left and right states partitioned
at the center of the grid. Additionally all three are run with the same
$\Gamma=4/3$ ideal gas equation of state. Using subscripts ``L'' and ``R'' to
denote left and right states, the initial data are:
$\rho_L=1$,     $P_L=1000$, $B^x_L=1$, $B^y_L=0$, 
$\rho_R=0.1$,   $P_R=1$,    $B^x_R=1$, $B^y_R=0$ for case MHDST-1;
$\rho_L=1$,     $P_L=30$,   $B^x_L=0$, $B^y_L=20$, 
$\rho_R=0.1$,   $P_R=1$,    $B^x_R=0$, $B^y_R=0$ for case MHDST-2; and
$\rho_L=1$,     $P_L=1$,    $B^x_L=0$, $B^y_L=1$, 
$\rho_R=0.125$, $P_R=0.1$,  $B^x_R=0$, $B^y_R=-1$ for the Brio-Wu case MHDST-3.

\begin{figure}
\includegraphics[width=0.48\textwidth]{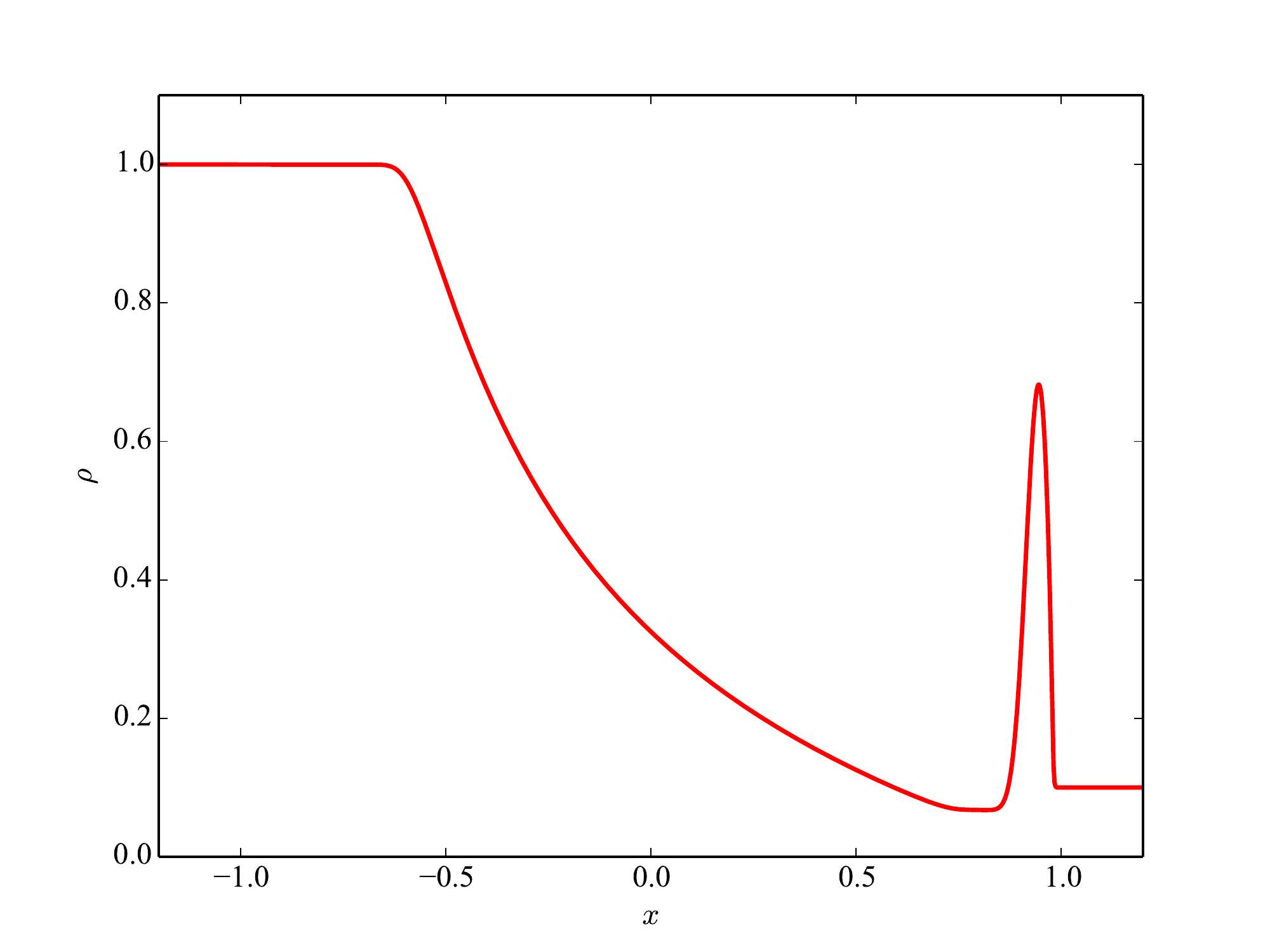} \\
\includegraphics[width=0.48\textwidth]{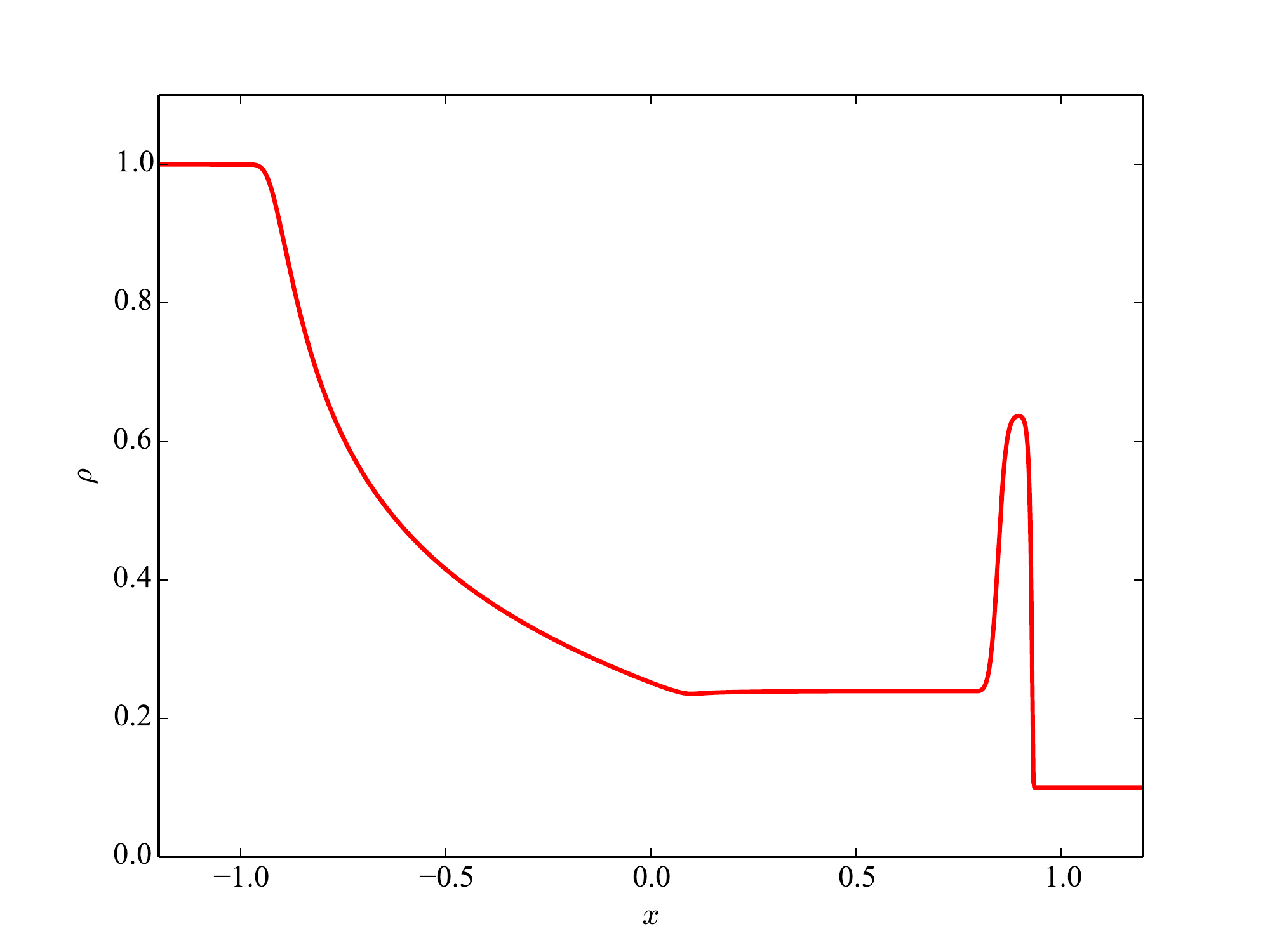} \\
\includegraphics[width=0.48\textwidth]{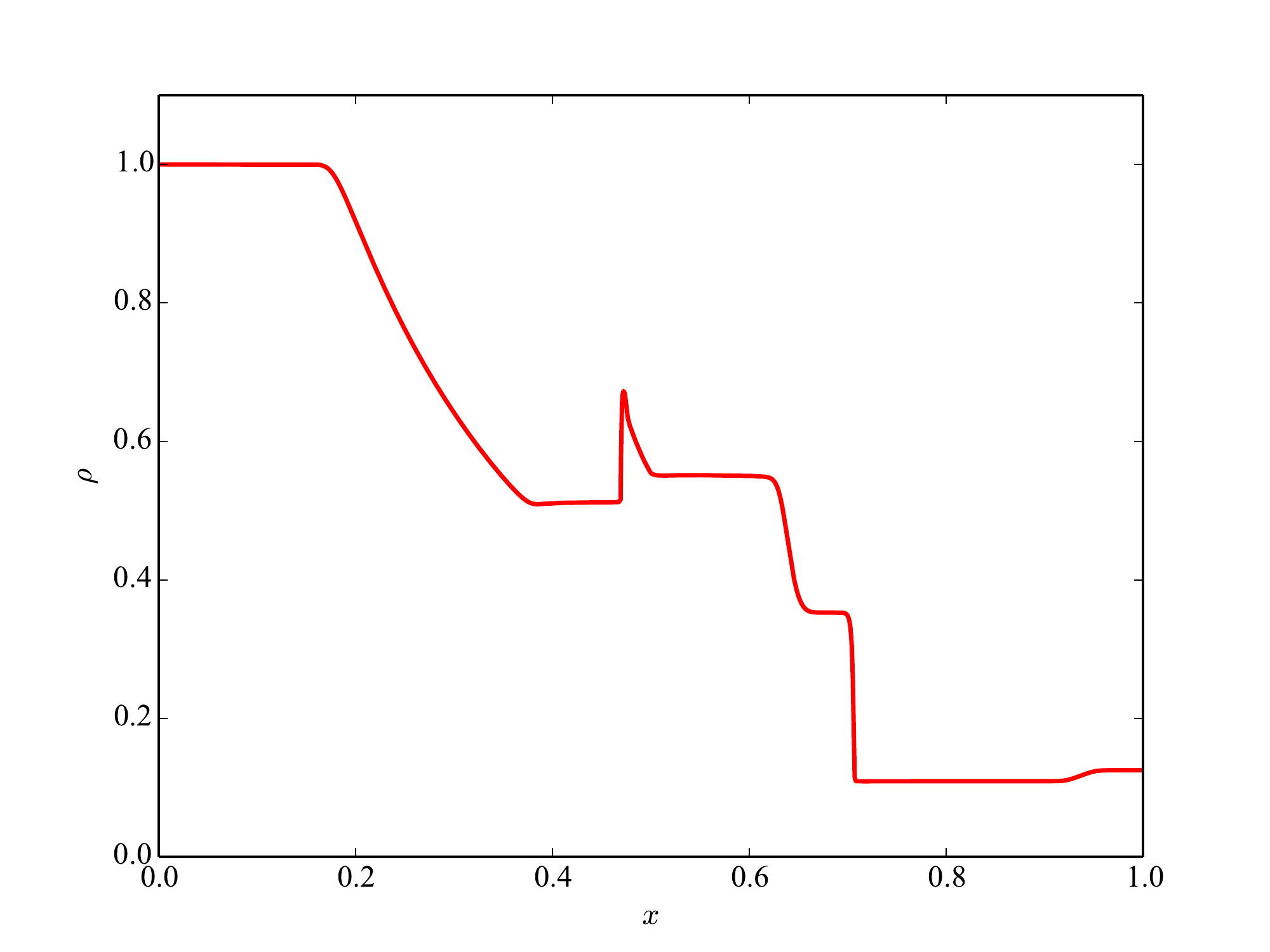}
\caption{Mass density $\rho$ for the three relativistic MHD shock tube tests using
artificial viscosity shock regularization and 1024 zones.
Solutions are plotted at times $t=1$, 1, and 0.5 for the 
MHDST-1, -2, and -3 cases, respectively.}
\label{fig:mhdtube}
\end{figure}

In Table \ref{tab:mhdtube}, we display the 
initial and final calculated values for each state of the shock tubes:
``Left'' is the initial left state, ``FL'' is the value at the foot
of the left fast rarefaction wave, ``SC'' is the value at the slow
compound wave, ``CDL'' is the left contact discontinuity,
``CDR'' is the right discontinuity, ``FR'' is the value at the foot
of the right fast rarefaction fan, and ``Right'' is the initial
right state. The numbers presented in Table \ref{tab:mhdtube} correspond
to calculations run with artificial viscosity, but we note that
results with slope limiting are similar and generally
match the viscosity results to within a few percent. Like the hydrodynamic shock tube tests, these problems were run
with flat boundary conditions, third order finite elements to demonstrate robustness of high
order DG on magnetized shock problems. Representative solutions of the
mass density calculated on a 1024 zone grid are plotted in Figure \ref{fig:mhdtube}
showing results from all three tests.

\begin{table}
  \centering
  \caption{Initial and final state solutions for the MHD shock tube tests
  \label{tab:mhdtube}}
  \begin{tabular}{l|ccccccccc}
    \tableline
    \tableline
              & Variable & Left  & FL     & SC   & CDL  & CDR  & FR   & Right  \\
    \tableline
    MHDST-1:  & $\rho$   & 1.0   & 0.07   & ...  & 0.69 & ...  & ...  & 0.1 \\
    MHDST-1:  & $P$      & 1000  & 28.5   & ...  & ...  & ...  & ...  & 1.0   \\
    \tableline
    MHDST-2:  & $\rho$   & 1.0   & 0.24   & ...  & 0.63 & ...  & ...  & 0.1 \\
    MHDST-2:  & $P$      & 30.   & 4.6    & ...  & 15.5 & ...  & ...  & 1.0   \\
    \tableline
    MHDST-3:  & $\rho$   & 1.0   & 0.51   & 0.67 & 0.55 & 0.35 & 0.11 & 0.125 \\
    MHDST-3:  & $P$      & 1.0   & 0.41   & 0.59 & 0.45 & 0.45 & 0.08 & 0.1   \\
    \tableline
  \end{tabular}
\end{table}

\subsection{Orszag-Tang}

\begin{figure}
\includegraphics[width=0.48\textwidth]{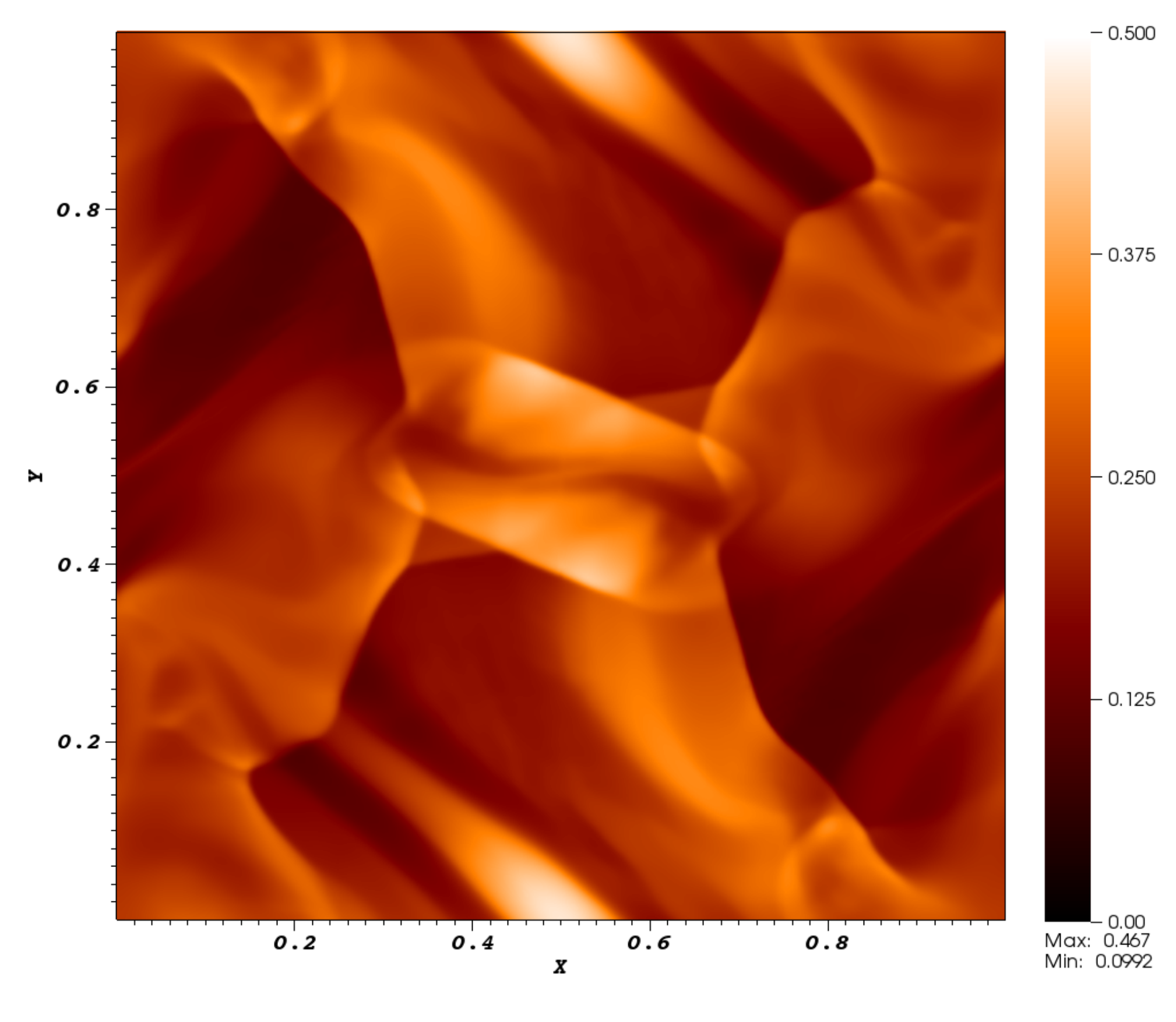}
\includegraphics[width=0.48\textwidth]{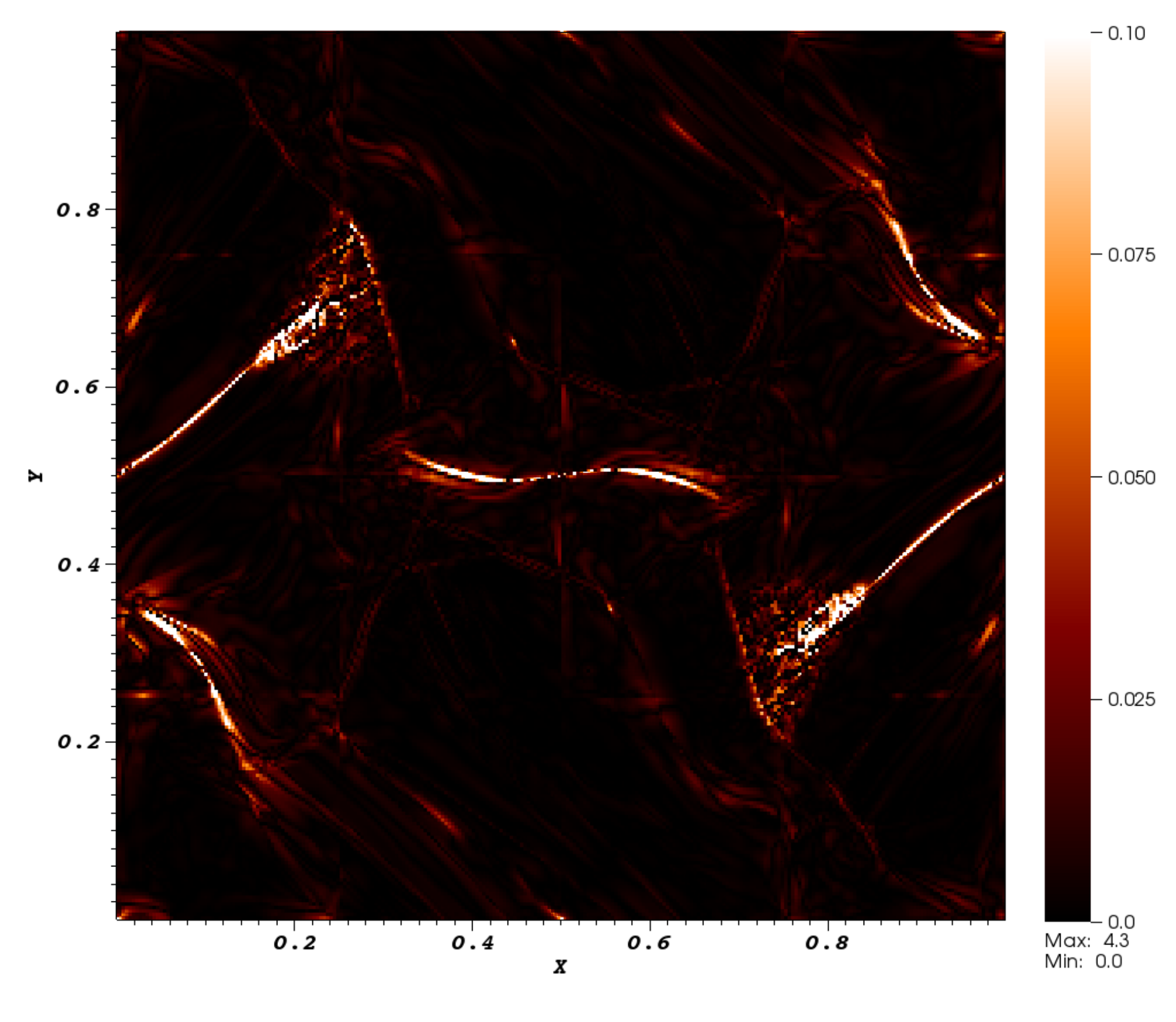}
\caption{Mass density (left) and divergence error (right) in the Orszag-Tang vortex test.
Images are shown at the final time $t=50$ calculated on a $256\times256$ grid
using third order finite elements.}
\label{fig:orszagimage}
\end{figure}

\begin{figure}
\includegraphics[width=0.6\textwidth]{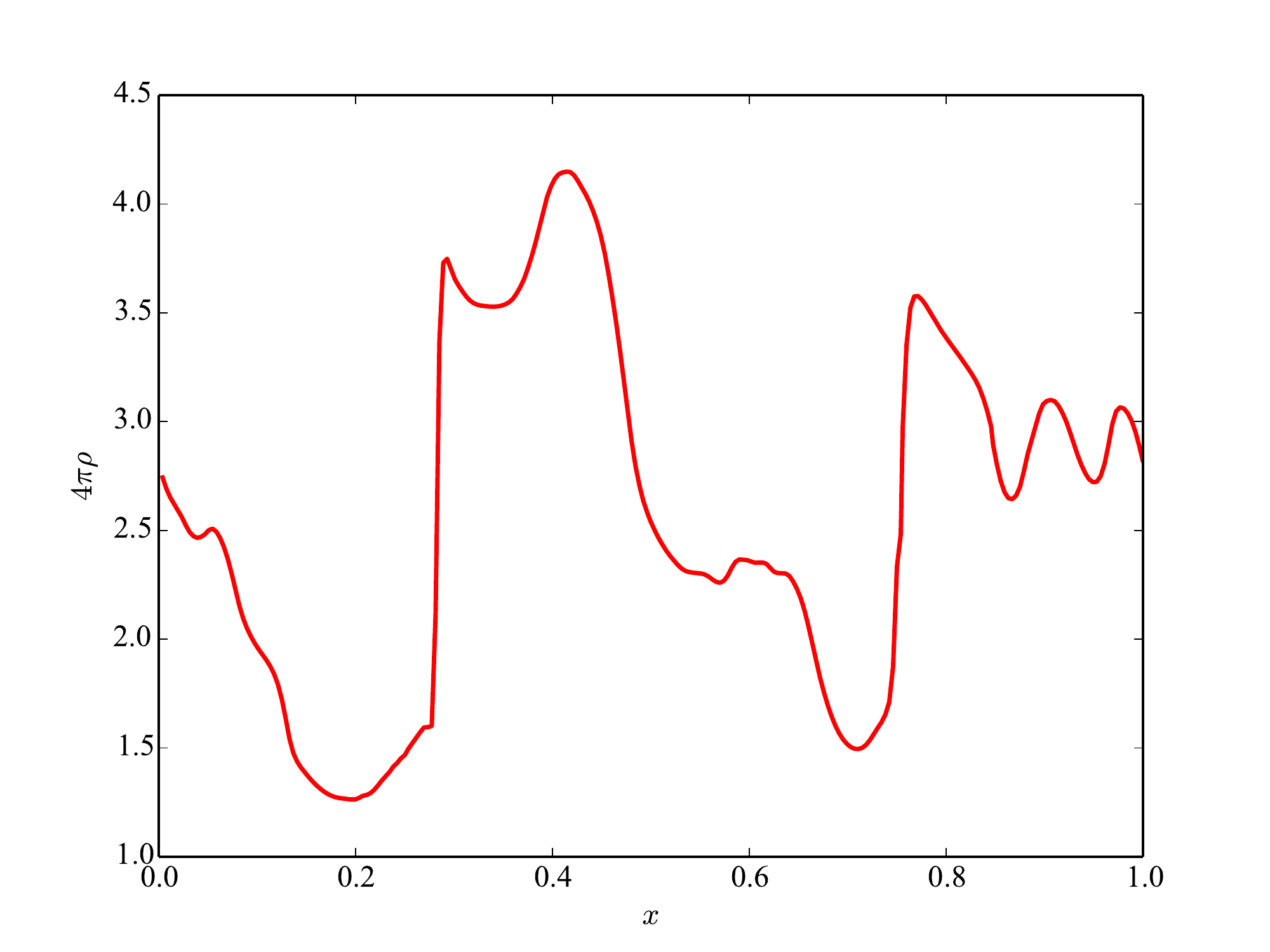}
\caption{Mass density multiplied by $4\pi$ as extracted along the 
horizontal line $y=0.75$ from Figure \ref{fig:orszagimage}.}
\label{fig:orszagslice}
\end{figure}

The Orszag-Tang vortex problem \citep{Orszag79} has become a standard
test of magnetic fields and divergence conservation. Numerous solutions
exist in the literature to which we can compare our results. In particular
we follow and adopt initial data from \citet{Sadowski14}
and \citet{Mocz14}: uniform density
$\rho=\Gamma^2/(4\pi)$, pressure $P=\Gamma/(4\pi)/{\cal C}^2$,
velocity $V^i=[-\sin(2\pi y), ~\sin(2\pi x), ~0]/{\cal C}$,
magnetic field $B^i=[-\sin(2\pi y), ~\sin(4\pi x), ~0]/\sqrt{4\pi}/{\cal C}$,
adiabatic index $\Gamma=5/3$, and a scale factor ${\cal C}=100$.
The problem is evolved out to a time of $t=50$, on a 
$256\times256$ unit two dimensional grid
$0 \le (x,y) \le 1$ with periodic boundary conditions applied in both directions.
Figure \ref{fig:orszagimage} shows the mass density and divergence
error at the final time using third order DG(2) finite elements. 
Also shown in Figure \ref{fig:orszagslice} is
a horizontal line out of the density multiplied by $4\pi$ 
along $y=0.75$, $4\pi \rho(x,y=0.75)$.
Both figures can be compared
to the corresponding results from Figure 4 of \citet{Sadowski14}
and Figure 16 of \cite{Mocz14}. Agreement
is excellent considering the solutions in \citet{Sadowski14} 
and \citet{Mocz14} were calculated on greater
$640\times640$ and $512\times512$ resolution  grids respectively.
In addition, we find global normalized divergence errors
$|\partial_i B^i|\Delta \ell/\sqrt{P_B}$ comparable to those reported
by \citet{Mocz14}: roughly a few $\times 10^{-3}$ that plateau early
and remain constant through most of the simulation.

\subsection{Kelvin-Helmholtz Instability}

In this section we present convergence studies of the linear growth phase
of the two-dimensional magnetized Kelvin-Helmholtz instability (KHI).
Following interesting observations by \citet{Mignone09} on the performance
of various Riemann solvers in this class of problems, \citet{Beckwith11}
published a brief study of KHI turbulence that provides a useful test of high
order adaptive numerical methods, so we 
attempt to duplicate some of their findings here. The problem
consists initially of two oppositely traveling $\Gamma=4/3$ fluids
in pressure equilibrium $P=1$, densities $\rho=1$ and $10^{-2}$, and the following velocity profiles:
\begin{equation}
  V^x = -V_\mathrm{shear} \tanh\left(\frac{y-y_0}{a}\right)	~,
\end{equation}
\begin{equation}
  V^y = -A_0 V_\mathrm{shear} ~\sin(2\pi x) ~\exp\left(-\left(\frac{y-y_0}{\sigma}\right)^2\right)	~,
  \label{eqn:KHVy}
\end{equation}
with shear velocity $V_{shear}= 0.5$, perturbation amplitude $A_0=0.1$, shear layer
thickness $a=0.01$, characteristic length scale $\sigma=0.1$, and interface
position $y_0= L_y/2$ where $L_y$ is the grid length along the $y$ axis. The density
is linearly interpolated similar to the shear velocity $V^x$ profile along the $y$-axis
so that $\rho=1$ in regions with $V^x=0.5$ and smoothly extended to $\rho=10^{-2}$ in regions
with $V^x=-0.5$. A single component magnetic field is introduced aligned along the $x$-direction
with $B^x=10^{-3}$. In addition, one percent Gaussian perturbations are applied
to both $x$ and $y$ components of the velocity, modulated by the same exponential
damping function used in equation (\ref{eqn:KHVy}).
The computation domain covers $0 \le (x,y) \le 1$, and for the AMR calculations
is resolved with a base grid of $64\times64$ cells. Periodic (reflection)
boundary conditions are enforced along the $x$($y$)-axis.
Three or four additional levels of mesh refinement are applied to capture
the fluid interface at effective $512\times512$ 
or $1024\times1024$ resolutions, adaptively refining
(and de-refining) on the dimensionless slope of the mass density.
The threshold refinement (derefinement) criteria for all cases is set to $s_r$ = 0.05 (0.001),
where $s_r = |\ell^i \partial_i\rho|/\overline{\rho}$, $\ell^i$ is a vector of
cell widths in each spatial dimension, and $\overline{\rho}$ is the local
(nearest neighbor) average of the mass density.

\begin{figure}
\includegraphics[width=0.6\textwidth]{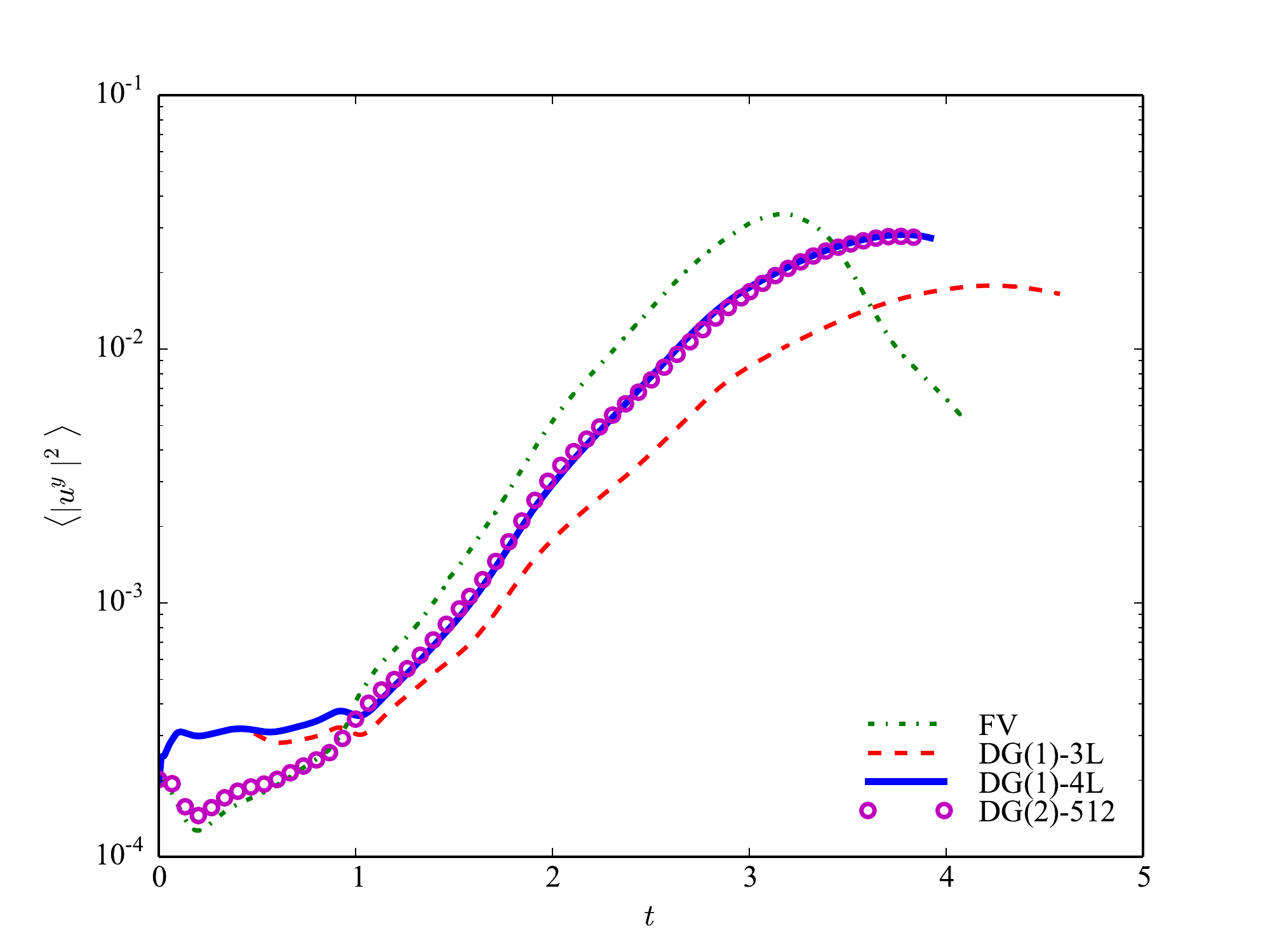}
\caption{Volume averaged transverse four-velocity $\langle|u^y|^2\rangle$ 
as a function of time over the linear growth phase of the 
magnetized Kelvin-Helmholtz instability.
The dot-dashed line corresponds to a converged finite volume
solution (case KHFV) resembling very closely the results from \citet{Beckwith11},
including slope, magnitude, and peak location. The remaining curves represent
the DG(1) solution with 3 AMR levels (case KHDG1-3L, dashed line), 
DG(1) with 4 AMR levels (case KHDG1-4L, solid line),
and DG(2) on a single $512\times512$ grid (run KHDG2-512, circles).}
\label{fig:KHUy}
\end{figure}

A good diagnostic of the linear growth stage of the KHI is
the temporal history of the square of the transverse four-velocity 
weighted by cell volume and averaged over the entire grid, $<|u^y|^2>$.
This quantity is plotted in
Figure \ref{fig:KHUy} for four cases: KHFV representing the converged second order
finite volume solution, KHDG1-3L using second order DG(1) with 3 AMR levels,
KHDG1-4L using DG(1) with 4 AMR levels, and
KHDG2-512 using third order DG(2) (both adaptive and fixed polynomial order) on a uniform $512\times512$ mesh. 
The finite volume calculation (KHFV, solid line) reproduces
converged results from \citet{Beckwith11}, matching the slope,
magnitude, and peak position in time. These four calculations collectively demonstrate
convergence towards the resolved solution with both mesh and basis order refinement.
Notice in particular that run KHDG1-4L (second order with 4 AMR levels, or
effectively $1024\times1024$ resolution across the interface) is nearly identical to
the result of KHDG2-512 (third order with $512\times512$ resolution),
and that both significantly improve compared
to the lowest resolution result KHDG1-3L (second order with 3 AMR levels).

\begin{figure}
\includegraphics[width=0.8\textwidth]{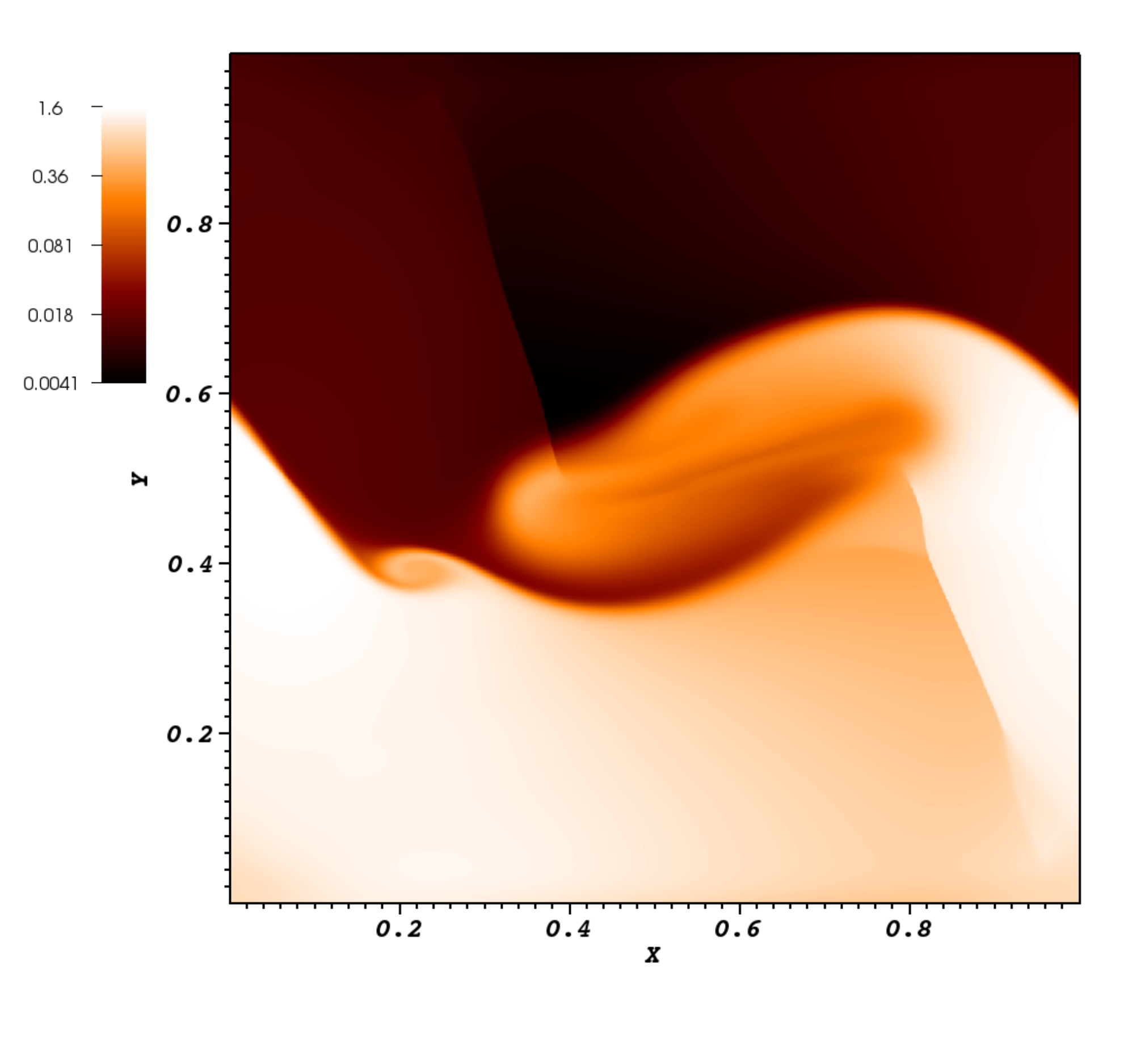}
\caption{Log of the mass density in the magnetized Kelvin-Hemholtz instability test shown
at time $t\approx3.3$ using the DG(1) method on a single
$1024\times1024$ grid.}
\label{fig:KHimage}
\end{figure}

The density distribution is shown in Figure \ref{fig:KHimage}
at time $t\approx 3.3$ using the DG(1) method on a
single $1024\times1024$ grid.
Interestingly, this calculation exhibits signs of a developing
secondary vortex at $x\approx 0.25$ that is not present in second
order calculations with resolutions less than $1024\times1024$.
Similar features however are observed at lower resolutions provided
high order (greater than second) finite element representations are utilized.
The third order KHDG2-512 calculation, for example, produces a similar feature at
$512\times 512$ resolution.
Sensitivities associated with the development of a
secondary vortex have been observed by \citet{Mignone09} 
and \citet{Beckwith11} who attributed this behavior to the accuracy of the Riemann
solver and its ability to capture the contact discontinuity. That
DG evolves this feature with a 2-speed HLL Riemann solver with no
contact steepening is encouraging and represents yet another potential benefit
of the DG methodology.

\subsection{Bondi Accretion}

A popular test with general relativistic spacetime curvature source terms
is radial accretion onto a compact Schwarzschild
black hole. The analytic solution is characterized by a critical
point $r_c$ in the flow \citep{Michel72}
\begin{equation}
(u_c^r)^2 = \frac{M}{2r_c} 	~,
\end{equation}
\begin{equation}
v_c^2 = \frac{(u_c^r)^2}{1-3(u_c^r)^2} = \frac{(1+n)T_c}{n(1+(1+n)T_c)} ~,
\end{equation}
where $u_c^r$ and $v_c$ are the radial 4-velocity and sound speed at
the critical point, respectively, $M$ is the black hole mass,
$n=1/(\Gamma-1)$ is the polytropic index, and $T=P/\rho$ is the fluid temperature.
The solution is completed with the following parametrization
\begin{equation}
T^n u^r r^2 = C_1	~,
\end{equation}
\begin{equation}
(1+(1+n)T)^2 ~\left(1 - \frac{2M}{r} + (u^r)^2\right) = C_2	~.
\end{equation}
The constants $C_1$ and $C_2$ are fixed by choosing the critical radius
$r_c=8GM/c^2$, setting $\Gamma=4/3$, and defining the mass density at the 
critical radius ($\rho_c$) by setting the mass accretion rate to 
$\dot{M}=4\pi r_c^2\rho_c u^r_c = -1$.

\begin{figure}
\includegraphics[width=0.6\textwidth]{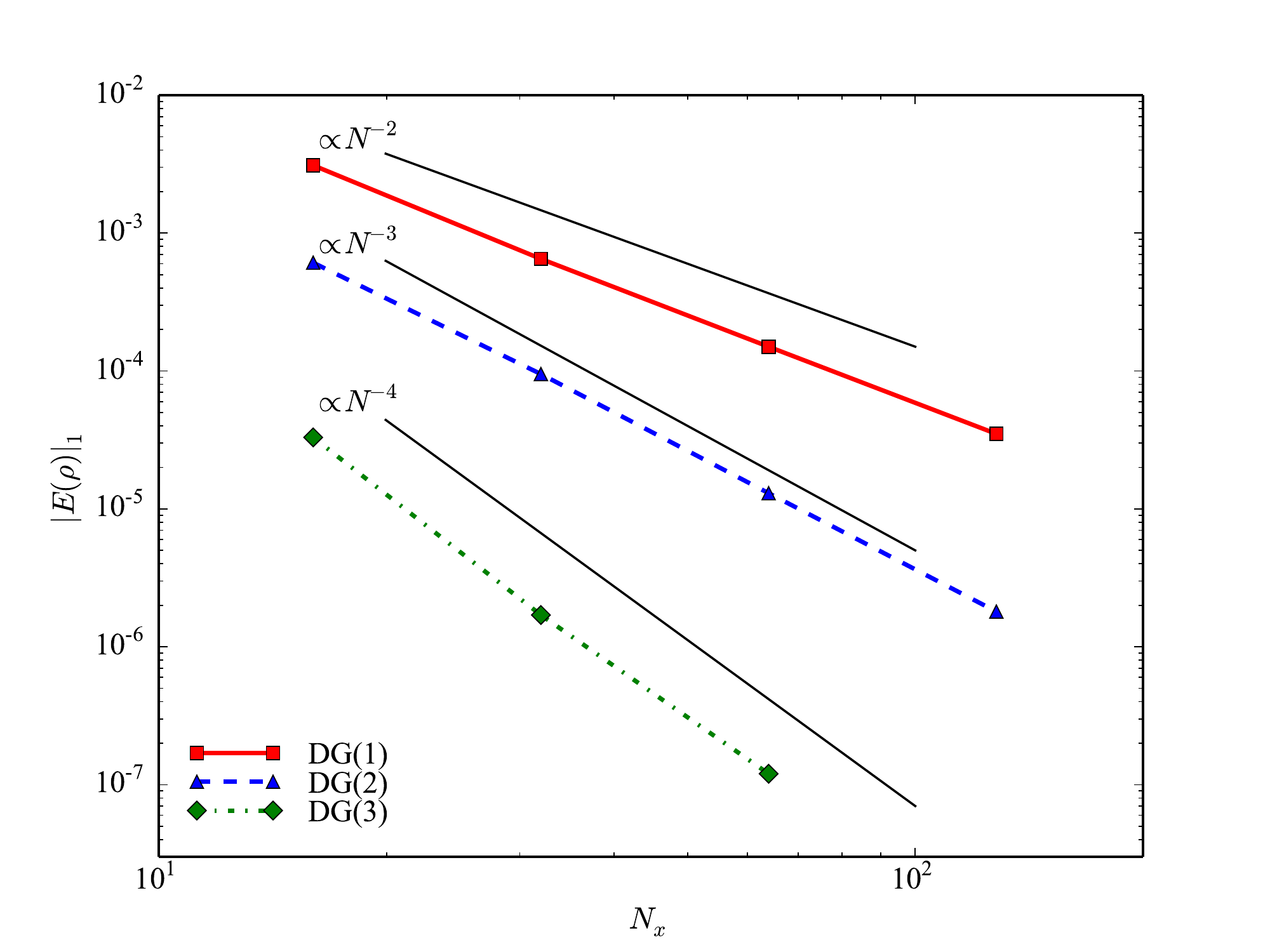}
\caption{Plot of the L1-norm errors in $\rho$ for the Bondi accretion test.}
\label{fig:bondi_error}
\end{figure}

We choose spherical Kerr-Schild coordinates for this test to cover
a two-dimensional computational domain bounded in radius from
$r=0.98 r_{BH}$ to $r=20 GM/c^2$,
where $r_{BH}=2GM/c^2$ is the radius of the black hole horizon.
The angular extent is a thin wedge centered along the equatorial
symmetry axis of width $\Delta\theta = \pi/20$.
The Bondi solution is initialized from the outset at $t=0$ then evolved
over a time interval of $5 GM/c^2$. Constant boundary conditions consisting
of the analytic solution are imposed at the inner and outer radial boundaries, 
and symmetric boundaries are enforced in the angular coordinate.
Accuracy and convergence are evaluated by calculating L1-norm errors in density
between initial and final times along the equatorial
plane over the entire radial extent of the grid.
Although {\sc CosmosDG} supplies analytic metric gradients for many 
black hole metric representations, for this test we instead evaluate gradients
numerically in order to test our implementation of finite element gradient operators.
A series of nine calculations were performed with three DG orders (2nd, 3rd, 4th)
and three grid resolutions:
$N_r\times N_\theta =$ $16\times2$, $32\times4$, $64\times8$,
where $N_r$ and $N_\theta$ are the number of zones along the radial
and angular directions.
All calculations were run with 4th order time integration using the 5-stage Runge-Kutta method.
Like previous smooth field tests, we find DG methods
produce significantly smaller evolution errors than FV, an
order of magnitude or more depending on the basis order, and converge
to the analytic solution at the appropriate rate. 
For example, FV produces L1-norm errors of $3.8\times10^{-4}$ in
mass density computed on a $64\times8$ grid.
Equivalent errors from DG(1), DG(2) and DG(3) methods
come to $1.5\times10^{-4}$, $1.3\times10^{-5}$  and $1.2\times10^{-7}$ respectively.
In addition we find with each doubling of zones, errors are reduced
by factors of about $2^{p+1}$ for methods DG($p$) as expected and as
demonstrated in Table \ref{tab:bondi_error} and Figure \ref{fig:bondi_error}.

This hydrodynamic black hole accretion test can be adapted to include
a radial magnetic field satisfying $\partial_r B^r=0$ and which 
does not alter the analytic solution for any of the primitive fields
($\rho$, $p$ or $u^r$). Although this treatment does not satisfy the
full Maxwell equations \citep{Anton06}, it is a useful nontrivial
test of magnetic fields in the code. 
We set the magnitude of the magnetic field by
$|b|^2/\rho = 10.56$ at the critical point, effectively
equating hydrodynamic and magnetic pressures at $r=r_c$.
Performing equivalent calculations (identical grids, resolutions, fluid and black
hole parameters) as the hydrodynamic version, we find errors
similar to those presented in Table \ref{tab:bondi_error}.

\begin{table}
  \centering
  \caption{L$1$-norm errors in $\rho$ for the Bondi accretion test
  \label{tab:bondi_error}}
  \begin{tabular}{l|cccc}
    \tableline
    \tableline
            & $16\times2$        & $32\times4$        & $64\times8$         \\
    \tableline
    DG(1)   & $3.1\times10^{-3}$ & $6.5\times10^{-4}$ & $1.5\times10^{-4}$  \\
    DG(2)   & $6.1\times10^{-4}$ & $9.5\times10^{-5}$ & $1.3\times10^{-5}$  \\
    DG(3)   & $3.3\times10^{-5}$ & $1.7\times10^{-6}$ & $1.2\times10^{-7}$  \\
    \tableline
  \end{tabular}
\end{table}

\subsection{Magnetized Black Hole Torus}

For a final test we expand on the Bondi accretion problem and consider a magnetized
torus of gas orbiting around a rotating black hole. There is no analytic solution
for this problem so in its place we instead compare a DG finite element solution
against a comparable finite volume calculation.
Like the Bondi test we use spherical Kerr-Schild 
spacetime coordinates, but set the spin of the black hole to $a/M=0.9$,
the specific angular momentum of the torus to $\ell/M=4.3$, and the
surface potential to $(u_t)_{in} = -0.98$, producing a torus of mass 
$M_{torus} = 1390 M$ orbiting
in rotational equilibrium around the central potential of the black hole.
The background regions are initially set to be
cold ($e=10^{-6} e_{\text{max}}$), low density ($\rho=10^{-6}\rho_{\text{max}}$),
and static ($V^i=0$), where $e_{\text{max}}$ and $\rho_{\text{max}}$ are the
internal energy density and mass density at the pressure maximum of the torus
located at radius $r_{\text{cent}}=15.3 GM/c^2$.
The orbital period at $r_{\text{cent}}$ is $t_{orb} = 380 GM/c^3$.
The torus is set up with a polytropic constant $\kappa = 0.01$ and
(along with the background gas) obeys an ideal
gas equation of state with adiabatic index $\Gamma=5/3$.

In order to seed
the magnetorotational instability (MRI), the torus is threaded initially with
a weak poloidal magnetic field derived from the following vector potential:
\begin{equation}
A_\phi =
  \begin{cases}
      k (\rho - \rho_{\text{cut}}) & \text{if}\quad \rho > \rho_{\text{cut}} ~, \\
      0                            & \text{otherwise} ~,
  \end{cases}
\end{equation}
where $\rho_{\text{cut}}=0.5\rho_{\text{max}}$ effectively keeps the
magnetic field inside the torus surface. The field is normalized
by defining the constant $k$ so that $\beta=P/P_B \ge 2$ throughout the torus.
Although initially confined to the torus, eventually the field evolves to
fill most of the background with a magnetized corona and high-$\beta$ outflows, in
addition to seeding the MRI and launching an accretion flow from the torus to the
black hole. It is this accretion flow that we use as a diagnostic for comparing
results from the different numerical methods.

This problem is run on a two-dimensional, azimuthally symmetric grid
resolved effectively with $128\times128$ cells: The finite volume calculation
representing the ``known'' solution is run on a single $128\times128$ grid, while
the finite element calculations on a 2-level nested grid with $64\times64$ 
base resolution to demonstrate simultaneous use of hierarchical mesh 
and basis order refinement. We typically simulate accretion disks
in this mode (with nested grids) to achieve greater resolution along the equatorial
plane while simultaneously avoiding refining along the pole-axis to relax the Courant constraint.
However, we have verified that replacing nested grids with
fully adaptive mesh refinement produces similar results.
We also use a logarithmic radial coordinate of
the form $\eta=1+\ln(r/r_{\text{BH}})$ and a concentrated angular coordinate
$x_2$ such that $\theta=x_2+\sin(2x_2)/4$ to further increase the resolution
near the black hole horizon and along the equator. The grid covers
$0.1\pi \le \theta \le 0.9\pi$ and $0.98r_{\text{BH}} \le r \le 120 M$,
resulting in cell widths of $\Delta r \approx 0.05 GM/c^2$
near the inner radial boundary and $\Delta r \approx 0.5 GM/c^2$ near the
initial pressure maximum of the torus. By comparison, the characteristic
wavelength of the MRI is $\lambda \equiv 2\pi v_A/\Omega \approx 2.5 GM/c^2$
near the initial maximum.
Symmetric (reflective) boundary conditions are imposed along the radial (angular) directions.

\begin{figure}
\includegraphics[width=\textwidth]{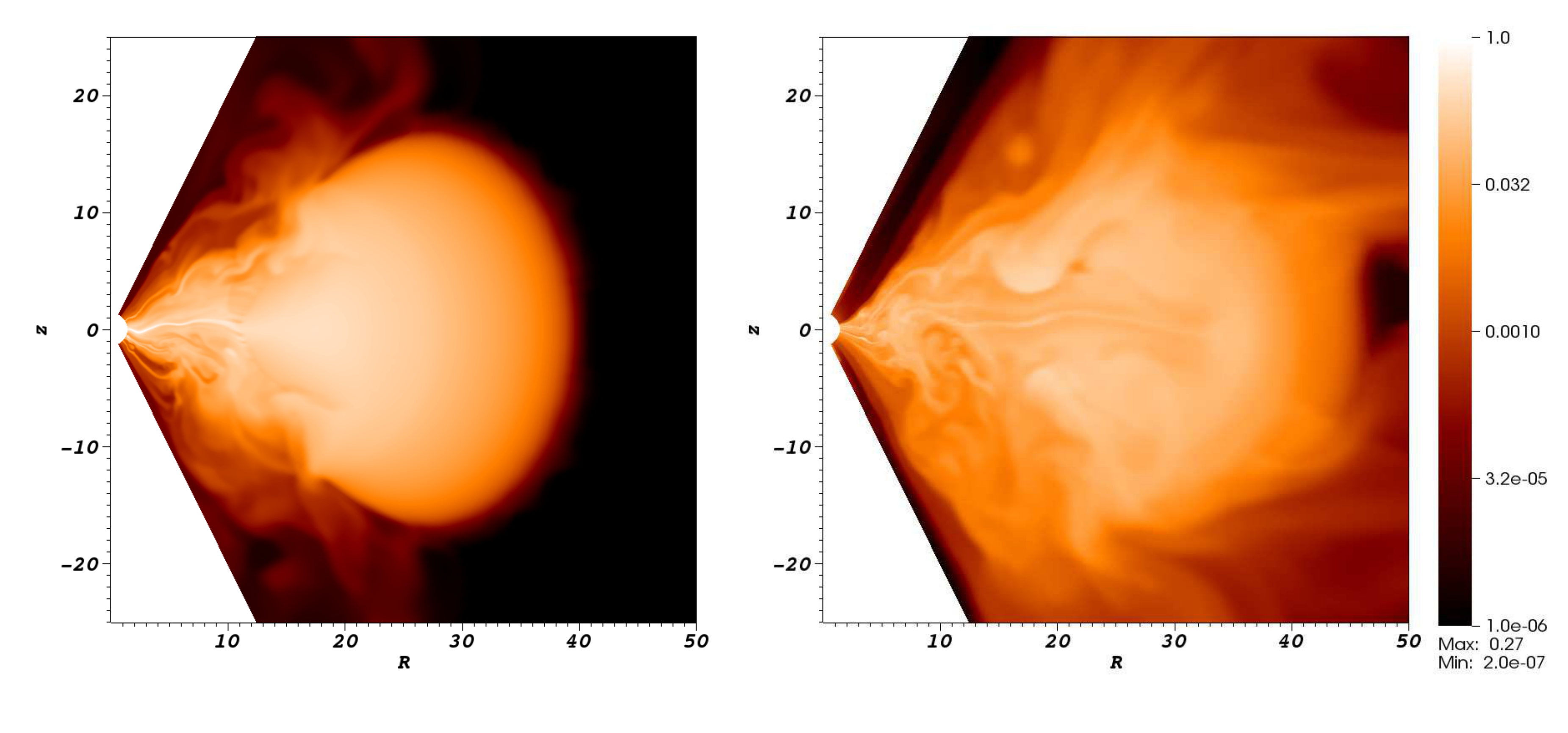}
\caption{Log of the mass density in the magnetized black hole torus accretion test 
calculated with the finite volume method. 
The left image corresponds to $t=199$ ($0.5~t_{orb}$), 
and the right to $t=760$ ($2~t_{orb}$).}
\label{fig:BHTimageFV}
\end{figure}

\begin{figure}
\includegraphics[width=\textwidth]{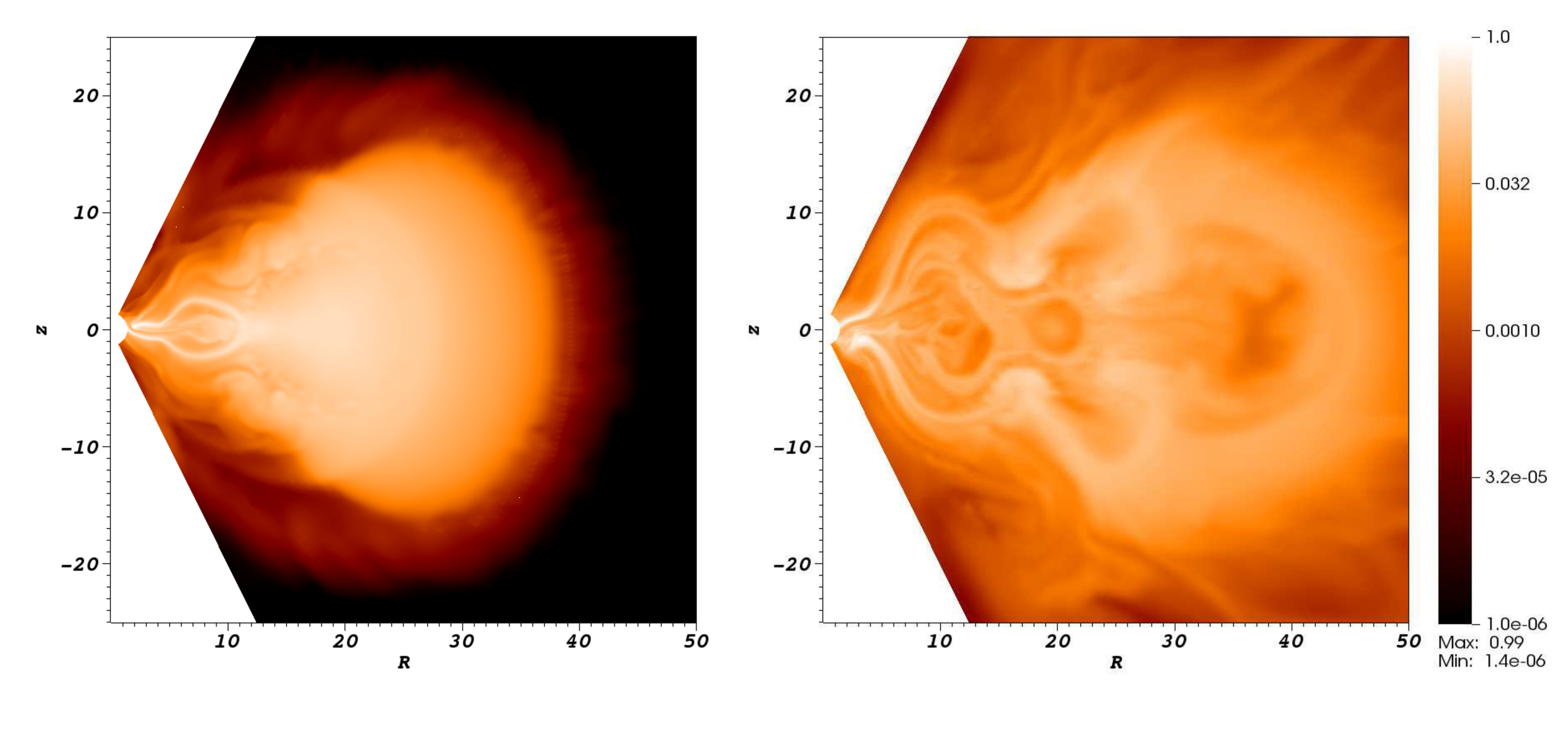}
\caption{Log of the mass density in the magnetized black hole torus accretion test 
calculated with the adaptive third order DG finite element method. 
The left image corresponds to $t=199$ ($0.5~t_{orb}$), 
the right to $t=760$ ($2~t_{orb}$).}
\label{fig:BHTimageDG}
\end{figure}

\begin{figure}
\includegraphics[width=0.8\textwidth]{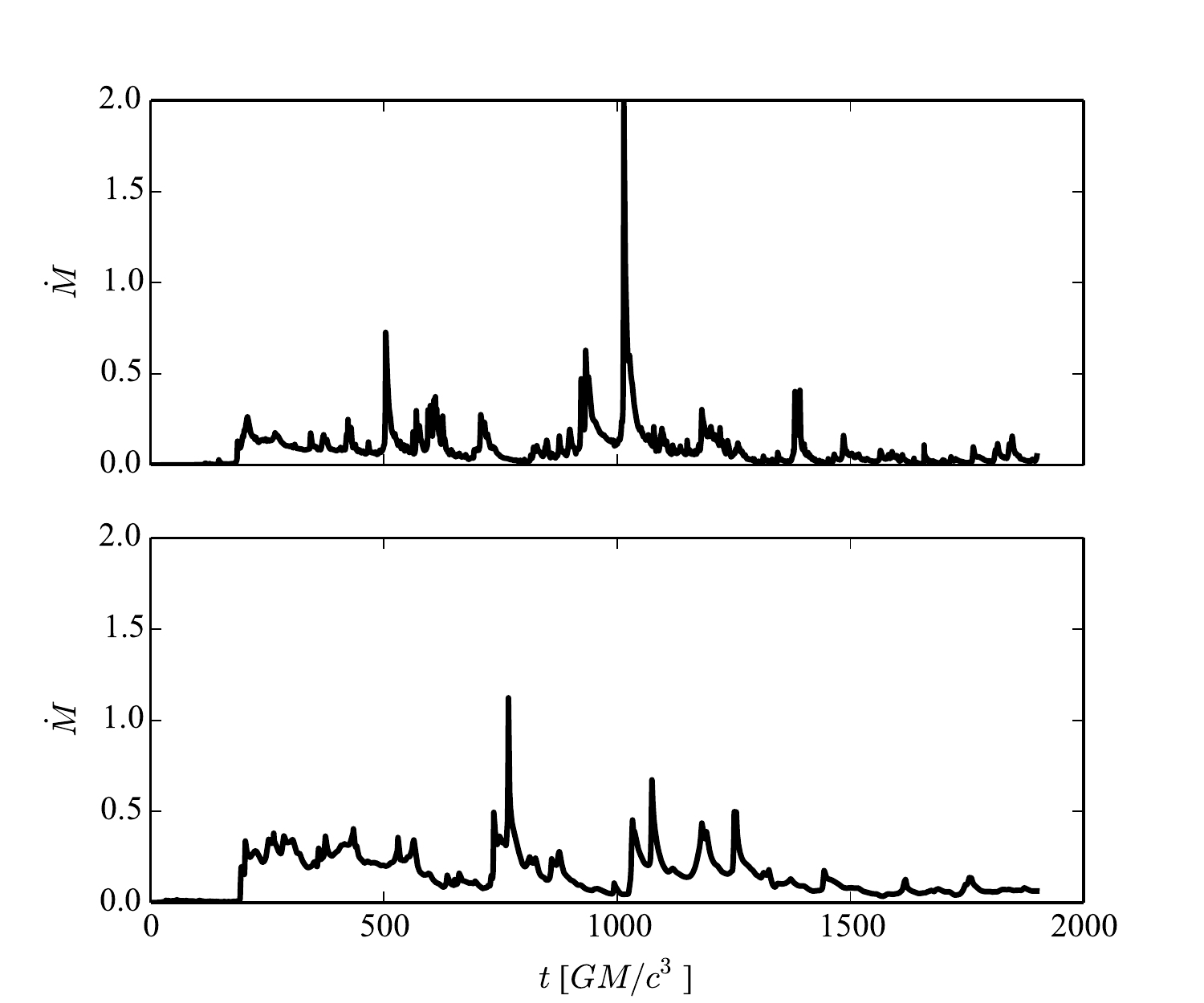}
\caption{Mass accretion rates $\dot{M}$ across the inner radial boundary,
comparing the finite volume (case BHT-FV, top) and
adaptive, third-order finite element (case BHT-DG2, bottom) solutions.}
\label{fig:BHTrate}
\end{figure}

Our comparison is between a traditional finite volume
with piecewise parabolic reconstruction (case BHT-FV)  
and an adaptive-order-refinement using second-order finite elements in the
background gas and third order in the torus body (case BHT-DG2), refining and derefining 
the polynomial order on the gas density (refining when the density exceeds 0.0005, derefining
when density drops below 0.0001). Figures \ref{fig:BHTimageFV}
and \ref{fig:BHTimageDG} show images of the logarithmic gas density,
comparing BHT-FV and BHT-DG2 solutions at two different times in the evolutions.
The left images correspond to an early time when the torus stream first hits the horizon.
The right images show a later snapshot when the flow has fully developed after
a couple of orbits.
Figure \ref{fig:BHTrate} compares mass accretion rates as a function of time for each of the cases.
We expect differences due to the turbulent nature of these calculations, but the level of differences between the methods appears significant, paying
particular attention to the temporal variability of the mass accretion,
the duration of the flow period before it begins to taper off, 
and the total accreted mass (about 12\% and 19\% of the initial
torus mass after five orbits for cases BHT-FV and BHT-DG2, respectively).
These results are consistent with previous similar calculations \citep{Anninos05},
though we note the slightly greater total accreted mass for the finite
element calculation.
More exhaustive studies of black hole accretion will be conducted in future work
to understand better the accuracy and benefits of high order methods for capturing sub-zonal effects.
Here we have taken the first step towards this goal, validating
the DG finite element methodology and demonstrating its utility to this class of problems.

\section{Discussion}

We have developed a new version of {\sc Cosmos++}, called {\sc CosmosDG}, a code for both Newtonian and general relativistic
radiation magnetohydrodynamics, based on the Discontinuous Galerkin (DG) finite element formulation.
The code infrastructure in {\sc CosmosDG} was upgraded to accommodate
simultaneous use of cell-by-cell adaptive mesh refinement (AMR) 
together with cell-by-cell adaptive basis order refinement (AOR).
Our current implementation utilizes Lagrange interpolatory functions to construct
local finite element approximations of arbitrary spatial order, with multiple options 
for high order time integration, including 3rd order forward Euler, and 4th order
strong stability preserving Runge-Kutta.
Generalization to two and three dimensions is accomplished through
tensor products of one-dimensional basis functions.
Although our DG implementation can
construct basis polynomials of arbitrary order, in practice we find that 3rd or 4th order
is a practical upper  limit set by computational speed and saturation of errors 
due to machine precision constraints.

Two options are provided for the regularization of shock discontinuities:
artificial viscosity and/or slope limiters (they can
be invoked separately or together).
Artificial viscosity is applied in a conservative manner using 
covariant Laplacian smoothing operators with properly
normalized entropy or relativistic enthalpy diagnostics to trigger the application of viscosity
locally when cell interface jumps or excessive sub-zonal heating are detected. The slope limiter option
is a modification of the traditional minmod limiter commonly used in finite volume
and finite difference codes, generalized here to
work in multi-dimensions and for arbitrary order finite elements
by projecting high order solutions to a low order basis using a least squares
method to compute slopes. Both approaches were demonstrated to
provide adequate stabilization of shocks, for all polynomial orders.
We note that artificial viscosity is implemented as sub-zonal dissipation
terms, preserving the high order nature of DG($p$) in each cell. Slope limiting, however,
is currently applied across zone interfaces like traditional
finite volume methods, effectively folding high order sub-grid data
into second order solutions before enforcing monotonicity. We are currently
investigating a number of approaches for extending slope limiting to
work directly on unfiltered sub-grid data.

We demonstrated the ability of DG finite element methods to 
achieve arbitrarily high order convergence on perturbation problems with
smooth profiles (i.e., sonic and magnetosonic waves), limited only by 
analytic solution and machine precision limits. 
Our extensive testing of DG methods furthermore proved them equal to or, in most cases, better than
finite volume methods, using high resolution shock capturing
or central difference schemes, even for problems with highly
relativistic shocks, regimes with strong discontinuities
where high order can break down and lead to unwanted Gibbs effects.
Perhaps more importantly we have subjected this methodology to the rigors of
multi-dimensional modeling of energetic astrophysical environments,
complex hydrodynamic instabilities, and strong spacetime curvature,
successfully demonstrating its application to Bondi-Hoyle and
MRI-induced black hole accretion.

\acknowledgements

This work was performed in part under the auspices of the 
U.S. Department of Energy by Lawrence Livermore National Laboratory 
under Contract DE-AC52-07NA27344.  This work used the Extreme Science and Engineering Discovery Environment (XSEDE), which is supported by National Science Foundation grant number ACI-1053575.   PCF and DN acknowledge support from National Science Foundation grants AST-1211230 and AST-1616185.  PCF was supported in part for this work by the National Science Foundation under grant NSF PHY11-25915.


\begin{thebibliography}{35}
\expandafter\ifx\csname natexlab\endcsname\relax\def\natexlab#1{#1}\fi

\bibitem[{{Anninos} \& {Fragile}(2003)}]{Anninos03a}
{Anninos}, P., \& {Fragile}, P.~C. 2003, \apjs, 144, 243

\bibitem[{{Anninos} {et~al.}(2003){Anninos}, {Fragile}, \&
  {Murray}}]{Anninos03b}
{Anninos}, P., {Fragile}, P.~C., \& {Murray}, S.~D. 2003, \apjs, 147, 177

\bibitem[{{Anninos} {et~al.}(2005){Anninos}, {Fragile}, \&
  {Salmonson}}]{Anninos05}
{Anninos}, P., {Fragile}, P.~C., \& {Salmonson}, J.~D. 2005, \apj, 635, 723

\bibitem[{{Ant{\'o}n} {et~al.}(2006){Ant{\'o}n}, {Zanotti}, {Miralles},
  {Mart{\'{\i}}}, {Ib{\'a}{\~n}ez}, {Font}, \& {Pons}}]{Anton06}
{Ant{\'o}n}, L., {Zanotti}, O., {Miralles}, J.~A., {Mart{\'{\i}}}, J.~M.,
  {Ib{\'a}{\~n}ez}, J.~M., {Font}, J.~A., \& {Pons}, J.~A. 2006, \apj, 637, 296

\bibitem[{{Babu\u{s}ka} {et~al.}(1986){Babu\u{s}ka}, {Zienkiewicz}, {Gago}, \&
  {de A. Oliveira}}]{Babuska86}
{Babu\u{s}ka}, I., {Zienkiewicz}, O.~C., {Gago}, J., \& {de A. Oliveira},
  E.~R., eds. 1986, {Accuracy Estimates and Adaptive Refinements in Finite
  Element Computations} (John Wiley and Sons, Chichester)

\bibitem[{{Beckwith} \& {Stone}(2011)}]{Beckwith11}
{Beckwith}, K., \& {Stone}, J.~M. 2011, \apjs, 193, 6

\bibitem[{{Cockburn} {et~al.}(1989){Cockburn}, {Lin}, \& {Shu}}]{Cockburn89}
{Cockburn}, B., {Lin}, S.-Y., \& {Shu}, C.-W. 1989, Journal of Computational
  Physics, 84, 90

\bibitem[{{Cockburn} \& {Shu}(1989)}]{Cockburn89b}
{Cockburn}, B., \& {Shu}, C.-W. 1989, Mathematics of Computation, 52, 411

\bibitem[{{Cockburn} \& {Shu}(1998)}]{Cockburn98}
---. 1998, Journal of Computational Physics, 141, 199

\bibitem[{{De Villiers} \& {Hawley}(2003)}]{DeVilliers03}
{De Villiers}, J.-P., \& {Hawley}, J.~F. 2003, \apj, 589, 458

\bibitem[{{Fragile} {et~al.}(2012){Fragile}, {Gillespie}, {Monahan},
  {Rodriguez}, \& {Anninos}}]{Fragile12}
{Fragile}, P.~C., {Gillespie}, A., {Monahan}, T., {Rodriguez}, M., \&
  {Anninos}, P. 2012, \apjs, 201, 9

\bibitem[{{Fragile} {et~al.}(2014){Fragile}, {Olejar}, \&
  {Anninos}}]{Fragile14}
{Fragile}, P.~C., {Olejar}, A., \& {Anninos}, P. 2014, \apj, 796, 22

\bibitem[{{Gottlieb} \& {Shu}(1998)}]{Gottlieb98}
{Gottlieb}, S., \& {Shu}, C.~W. 1998, Mathematics of Computation, 67, 73

\bibitem[{{Guermond} {et~al.}(2011){Guermond}, {Pasquetti}, \&
  {Popov}}]{Guermond11}
{Guermond}, J.-L., {Pasquetti}, R., \& {Popov}, B. 2011, Journal of
  Computational Physics, 230, 4248

\bibitem[{{Harten} {et~al.}(1983){Harten}, {Lax}, \& B.}]{Harten83}
{Harten}, A., {Lax}, P., \& B., v. 1983, SIAM Rev., 25, 35

\bibitem[{{Hartmann} \& {Houston}(2002)}]{Hartmann02}
{Hartmann}, R., \& {Houston}, P. 2002, Journal of Computational Physics, 183,
  508

\bibitem[{{Johnson} {et~al.}(1984){Johnson}, {Navert}, \&
  {Pitkaranta}}]{Johnson84}
{Johnson}, C., {Navert}, U., \& {Pitkaranta}, J. 1984, Computer Methods in
  Applied Mechanics and Engineering, 45, 285

\bibitem[{{Kershaw} {et~al.}(1995){Kershaw}, {Prasad}, \& {Shaw}}]{Kershaw95}
{Kershaw}, W., {Prasad}, Y.-X., \& {Shaw}, G. 1995, Technical Report
  UCRL-JC-122104, Lawrence Livermore Laboratory

\bibitem[{{Kidder} {et~al.}(2017){Kidder}, {Field}, {Foucart}, {Schnetter},
  {Teukolsky}, {Bohn}, {Deppe}, {Diener}, {H{\'e}bert}, {Lippuner}, {Miller},
  {Ott}, {Scheel}, \& {Vincent}}]{Kidder17}
{Kidder}, L.~E., {et~al.} 2017, Journal of Computational Physics, 335, 84

\bibitem[{{Komissarov}(1999)}]{Komissarov99}
{Komissarov}, S.~S. 1999, \mnras, 303, 343

\bibitem[{{Kuzmin} \& {Turek}(2004)}]{Kuzmin04}
{Kuzmin}, D., \& {Turek}, S. 2004, Journal of Computational Physics, 198, 131

\bibitem[{{Meier}(1999)}]{Meier99}
{Meier}, D.~L. 1999, \apj, 518, 788

\bibitem[{{Michel}(1972)}]{Michel72}
{Michel}, F.~C. 1972, \apss, 15, 153

\bibitem[{{Mignone} {et~al.}(2009){Mignone}, {Ugliano}, \& {Bodo}}]{Mignone09}
{Mignone}, A., {Ugliano}, M., \& {Bodo}, G. 2009, \mnras, 393, 1141

\bibitem[{{Mocz} {et~al.}(2014){Mocz}, {Vogelsberger}, {Sijacki}, {Pakmor}, \&
  {Hernquist}}]{Mocz14}
{Mocz}, P., {Vogelsberger}, M., {Sijacki}, D., {Pakmor}, R., \& {Hernquist}, L.
  2014, MNRAS, 437, 397

\bibitem[{{Noble} {et~al.}(2006){Noble}, {Gammie}, {McKinney}, \& {Del
  Zanna}}]{Noble06}
{Noble}, S.~C., {Gammie}, C.~F., {McKinney}, J.~C., \& {Del Zanna}, L. 2006,
  \apj, 641, 626

\bibitem[{{Orszag} \& {Tang}(1979)}]{Orszag79}
{Orszag}, S.~A., \& {Tang}, C.-M. 1979, Journal of Fluid Mechanics, 90, 129

\bibitem[{{Radice} \& {Rezzolla}(2011)}]{Radice11}
{Radice}, D., \& {Rezzolla}, L. 2011, \prd, 84, 024010

\bibitem[{{Reed} \& {Hill}(1973)}]{Reed73}
{Reed}, W., \& {Hill}, T. 1973, Technical Report LA-UR-73-479, Los Alamos
  Laboratory

\bibitem[{{S{\c a}dowski} {et~al.}(2014){S{\c a}dowski}, {Narayan}, {McKinney},
  \& {Tchekhovskoy}}]{Sadowski14}
{S{\c a}dowski}, A., {Narayan}, R., {McKinney}, J.~C., \& {Tchekhovskoy}, A.
  2014, MNRAS, 439, 503

\bibitem[{{Schaal} {et~al.}(2015){Schaal}, {Bauer}, {Chandrashekar}, {Pakmor},
  {Klingenberg}, \& {Springel}}]{Schaal15}
{Schaal}, K., {Bauer}, A., {Chandrashekar}, P., {Pakmor}, R., {Klingenberg},
  C., \& {Springel}, V. 2015, \mnras, 453, 4278

\bibitem[{{Schwab}(1999)}]{Schwab99}
{Schwab}, C. 1999, {P- And Hp- Finite Element Methods: Theory and Applications
  in Solid and Fluid Mechanics} (Numerical Mathematics and Scientific
  Computation, London)

\bibitem[{{Shu} \& {Osher}(1988)}]{Shu88}
{Shu}, C.-W., \& {Osher}, S. 1988, Journal of Computational Physics, 77, 439

\bibitem[{{Spiteri} \& {Ruuth}(2002)}]{Spiteri02}
{Spiteri}, R.~J., \& {Ruuth}, S.~J. 2002, SIAM Journal on Numerical Analysis,
  40, 469

\bibitem[{{Zanotti} {et~al.}(2015){Zanotti}, {Fambri}, \&
  {Dumbser}}]{Zanotti15}
{Zanotti}, O., {Fambri}, F., \& {Dumbser}, M. 2015, \mnras, 452, 3010

\end{thebibliography}
\end{document}